\documentclass[11pt]{article}

\usepackage[margin=1in]{geometry}
\usepackage{times}
\usepackage{verbatim}
\usepackage{graphicx}
\usepackage{paralist}
\usepackage{amsmath,amsthm,amssymb,amsfonts}
\usepackage{caption}
\usepackage{setspace}
\usepackage[colorlinks,linkcolor=black,bookmarksopen,
  bookmarksnumbered,citecolor=black,urlcolor=blue]{hyperref}
\graphicspath{{Figures/}} % to be commented out later

\usepackage{subcaption}

\usepackage{verbatim}

\newcommand{\R}{ {\mathbb R} }

\usepackage{color}
\newcommand{\add}[1]{\textcolor{blue}{#1}}
\usepackage[normalem]{ulem}

\definecolor{darkgrn}{rgb}{0, 0.8, 0}

\newcommand{\rank}{\operatorname{rank}}
\newcommand{\Pred}{\operatorname{Pred}}

\usepackage[ruled,vlined]{algorithm2e}

\theoremstyle{plain}
\newtheorem{theorem}{Theorem}[section]
\newtheorem{lemma}[theorem]{Lemma}

\theoremstyle{definition}
\newtheorem{definition}[theorem]{Definition}
\theoremstyle{remark}

\theoremstyle{claim}

\usepackage{gensymb}
\usepackage{mathrsfs}

\newcommand{\maxip}{\textsc{Max-IP}}
\newcommand{\maxipsp}{\textsc{Max-IP} }

\newcommand{\ip}{\textsc{IP}}
\newcommand{\ipsp}{\textsc{IP} }

\newcommand{\cP}{\mathcal{P}} % collection of interesting paths
\newcommand{\hyppox}{\texttt{Hyppo-X}}

\usepackage{lscape} 

\newcommand{\Sig}{\operatorname{Sig}}
\newcommand{\IScr}{{\cal I}}

\usepackage[ruled,vlined]{algorithm2e}

\usepackage{xr}

\begin{document}

\title{\hyppox: A Scalable Exploratory  Framework for Analyzing Complex Phenomics Data}

\author{
  Methun Kamruzzaman\footnote{School of Electrical Engineering and Computer Science, Washington State University, Pullman, WA, 99164, USA; \href{mailto:md.kamruzzaman@wsu.edu}{md.kamruzzaman@wsu.edu}} \hspace*{0.3in}
  Ananth Kalyanaraman\footnote{School of Electrical Engineering and Computer Science, Washington State University, Pullman, WA, 99164, USA; \href{mailto:ananth@wsu.edu}{ananth@wsu.edu}} \hspace*{0.3in}
  Bala Krishnamoorthy\footnote{Department of Mathematics and Statistics, Washington State University, Vancouver, WA, 98686, USA; \href{mailto:kbala@wsu.edu}{kbala@wsu.edu}} \\
  Stefan Hey\footnote{Department of Agronomy, Iowa State University, Ames, IA, 50011-1085, USA; \href{mailto:shey@iastate.edu}{shey@iastate.edu}} \hspace*{0.3in}
  Patrick S. Schnable\footnote{Department of Agronomy, Iowa State University,Ames, IA, 50011-1085, USA; \href{mailto:schnable@iastate.edu}{schnable@iastate.edu}}
}
\date{\relax}
\maketitle

\begin{abstract}
  Phenomics is an emerging branch of modern biology that uses high throughput phenotyping tools to capture multiple environmental and phenotypic traits, often at massive spatial and temporal scales.
  The resulting high dimensional data represent a treasure trove of information for providing an in-depth understanding of how multiple factors interact and contribute to the overall growth and behavior of different genotypes. 
  However, computational tools that can parse through such complex data and aid in extracting plausible hypotheses are currently lacking. 
  In this paper, we present \hyppox, a new algorithmic approach to visually explore complex phenomics data and in the process characterize the role of environment on phenotypic traits.
  We model the problem as one of unsupervised structure discovery, and use emerging principles from algebraic topology and graph theory for discovering higher-order structures of complex phenomics data.
  We present an open source software which has interactive visualization capabilities to facilitate data navigation and hypothesis formulation.
  We test and evaluate \hyppox{} on two real-world plant (maize) data sets.
  Our results demonstrate the ability of our approach to delineate divergent subpopulation-level behavior. 
  Notably, our approach shows how environmental factors could influence phenotypic behavior, and how that effect varies across different genotypes and different time scales. 
  %Our tool also has the ability to identify divergent subpopulations within a large population. 
  %The results presented in this paper show a promising application of topology and its role in hypothesis extraction from high-dimensional data sets.  
  To the best of our knowledge, this effort provides one of the first approaches to systematically formalize the problem of hypothesis extraction for phenomics data.
  Considering the infancy of the phenomics field, tools that help users explore complex data and extract plausible hypotheses in a data-guided manner will be critical to future advancements in the use of such data.
\end{abstract}


\section{Introduction}\label{sec:intro}
%\IEEEraisesectionheading{\section{Introduction}\label{sec:intro}}


%\IEEEPARstart{H}{igh-throughput}
High-throughput
  technologies are beginning to change the way we 
observe and measure the natural world. 
In medicine, physicians are using imaging and other
specialized sampling devices to keep a longitudinal log of patients'
drug/therapy response and other disease-related phenotypes.  
In agricultural biotechnology, phenotyping technologies such as cameras and LiDARs
are being used to measure physiological and morphological
features of crops %as it grows
in fields.  
%Alongside these technologies,
Further, advancements in genotyping
technologies (sequencing) have made it possible to
characterize and track genetic diversity and changes
at a high resolution,
and decode genetic markers that are key to performance traits.
Taken together, advancements in these technologies
are leading to
a rapid explosion of high-dimensional data, obtained from a variety of sources.

A distinctive feature of these inherently high-dimensional 
data sets is that their generation is motivated more based 
on the availability and easy access to high-throughput technology
as opposed to specific working hypotheses. 
While there are some broad high-level questions or research
themes that motivate the collection of data, the specific questions that relate
to testable hypothesis and eventual discoveries (e.g., what genetic
variations impact a physical trait, or how a combination of environmental
variables control a phenotype) are \emph{not} readily available {\it a
priori}.

Consider the case of plant phenomics
\cite{bilder_phenomics:_2009,houle_phenomics:_2010}.
Understanding how different crop varieties or \emph{genotypes} ($G$) interact with
\emph{environments} ($E$) to produce different varying performance traits
(\emph{phenotypes} ($P$)) is a fundamental goal of modern biology
($G\times E\rightarrow P$) \cite{aspb_unleashing_2013,martin_plant_2013}.
To address this fundamental albeit broad goal,
plant biologists and farmers
have started to widely deploy an array of high-throughput sensing
technologies that measure tens of crop phenotypic traits in the field
(e.g., crop height, growth characteristics, photosynthetic activity). 
These technologies, comprising mostly of camera and other
recording devices, generate a wealth of images (visual, infrared,
thermal) and time-lapse videos that represent a detailed set of
observations of a crop population as it develops over the course of 
the growing season.  Additionally, environmental sensors help in
collecting daily field measurements that represent the growth 
conditions. Furthermore, through the use of sophisticated genotyping
technologies, the genotypes of the different crop varieties are also
cataloged. 

\begin{figure}[!h]
\centering
\includegraphics[keepaspectratio=yes, width=\linewidth]{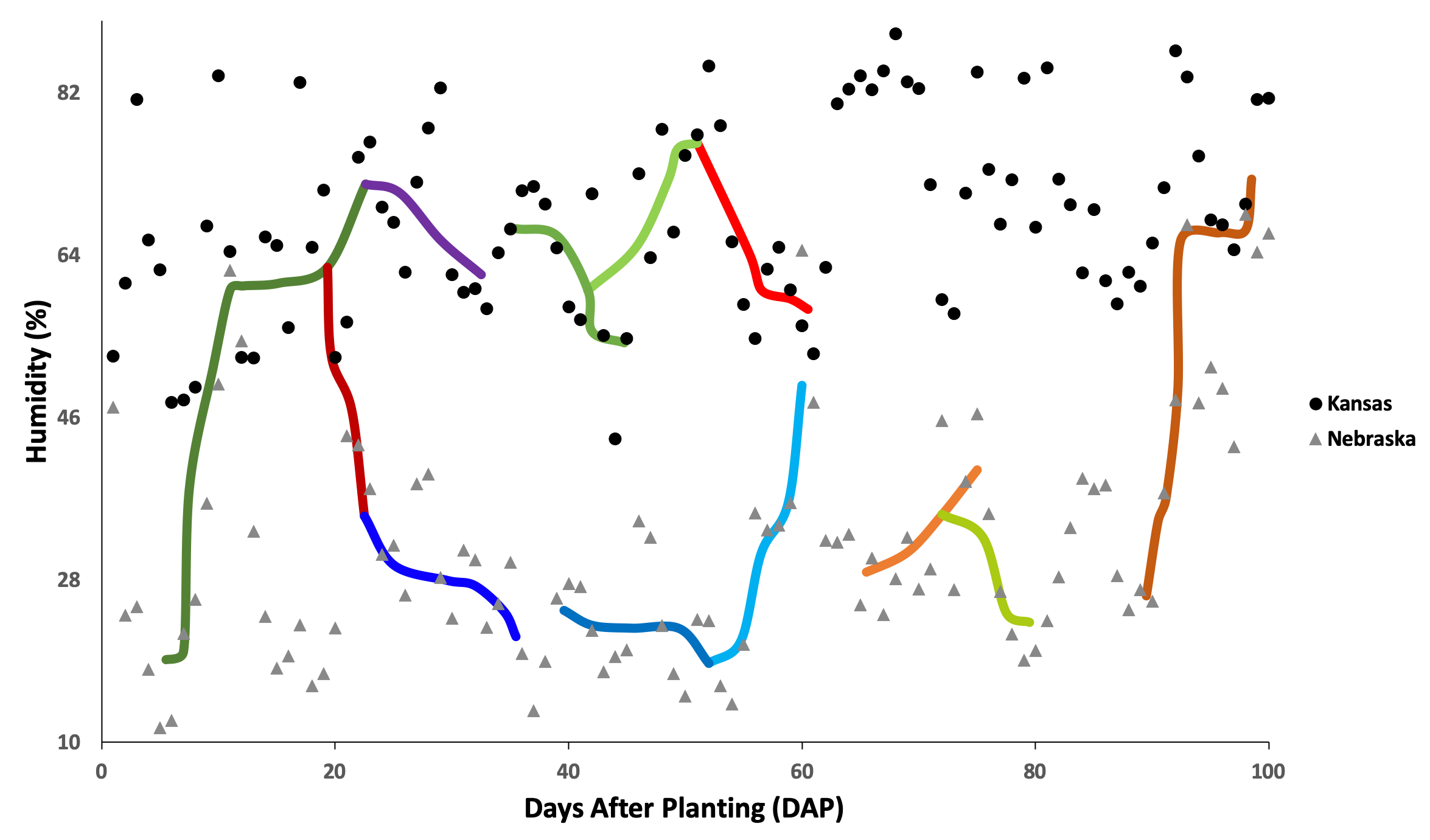}
\caption{
Scatter plot of a maize crop data set containing points grown in two
	locations---Kansas (KS) and Nebraska (NE). 
	Each data ``point'' is an [individual, date/time] combination, with
	x-axis representing the Days After Planting (time) and y-axis
	representing humidity as an environmental variable. 
	While the scatter plot shows higher humidity values in KS than in NE in
	general, it does not in itself have the capability to show
	intra-population variation with respect to a performance variable. 
	To this end, the ``interesting paths'' generated by our TDA framework
	can be useful. These paths are shown overlaid on the scatter plot; each
	path highlights a subset of points that are connected by consistent
	performance behavior (growth rate in this example). This could help us
	identify behaviorally-coherent subpopulations within large populations that are
	otherwise nontrivial to observe.
}
\label{fig:fig_2D_path}
\end{figure}

From this medley of plant genotypes, phenotypes, and environmental
measurements, scientists aim to extract plausible hypotheses that can be
field-tested and validated. However, the task remains significantly
challenging, mainly due to the dearth of automated software capabilities that are
capable of handling complex, high-dimensional data sets. Scatter plots 
(such as the example shown in \figurename~\ref{fig:fig_2D_path})
and correlation studies can reveal only high-level correlations and behavioral
patterns/differences within data. However, it is common knowledge that
different individuals or subgroups of individuals behave differently under
similar stimuli.  
For instance, while it is useful to know that a given
environmental variable (e.g., humidity) shows an overall positive
correlation to a performance trait (say, crop height), such high-level
correlations obfuscate the variations within a population---e.g., how
different subgroups or genotypes respond to different intervals in the environmental
values; or how one environmental variable interacts/interplays with
another to affect the performance trait; or how the same genotype expresses
variability in its performance under different environments (plasticity).

\subsection{Our Contributions}
\label{secContributions}

In this paper, we present a novel computational approach for extracting
hypotheses from high-dimensional data sets such as ones collected in phenomics. 
We formulate the problem of hypothesis
extraction as one of: (a) identifying the key connected structural
features of the given data, and (b) exploring the structural features
in a way to facilitate extraction of plausible hypotheses. 

\subsubsection{Structure Identification}
\label{struc-Iden}
Our approach uses emerging principles from \emph{algebraic topology}
as the basis to observe and discern structural features from
raw phenomics data. Algebraic topology is the field of mathematics 
dealing with the shape and connectivity of spaces \cite{Munkres1984,Ca2009}.  
There are multiple important properties of topology that make it
particularly effective for extracting structural features from
large, high-dimensional data sets. 
First, topology studies shapes in a {\em coordinate-free way}, which
enables comparison among data sets from diverse sources or
coordinate systems.  
Second, topological constructions are {\em not sensitive to small
changes in data}, and robust against noise.
Third, topology works with {\em compressed representations} of
spaces in the form of \emph{simplicial complexes} (or 
triangulations) \cite{Munkres1984}, which preserve information
relevant to how points are connected.  
Compared to more traditional techniques such as principal component
analysis, multidimensional scaling, and cluster
analysis, topological methods are known to be more sensitive to both
large and small scale patterns \cite{Lumetal2013}.

\subsubsection{Topological Object Exploration}
\label{topoObjExplo} 
While topological representations offer a compact way to represent and explore the data, 
the problem of how to navigate such representations in order to glean 
hypothesis information is still unexplored.
%\add{
In this paper, we formulate this problem formally as identifying a) \emph{interesting 
flares} and b) \emph{interesting paths}. 
The features we target encapsulate different properties of the data, as detailed below.
%}

%\add{
{\bf Interesting flares:}
Flares show how a subset of points (i.e., subpopulation) branches into
smaller subpopulations when exposed to certain environmental stimuli---e.g., a set of
plant individuals (or varieties) that shows divergent behavior in their growth
characteristics when one of the environmental parameters (say, temperature) crosses a certain value. 
Identifying such flares could help us identify subpopulations of interest and track
their behavioral evolution at a finer granularity of the population.
%}

%\add{
{\bf Interesting paths:} 
A \emph{path}, on the other hand, highlights a trail of point clusters along which a
``performance'' variable increases (or decreases). In other words, a path can reveal
different subpopulations that are prevalent in different performance intervals.
This can in turn help us contrast different population subtypes or subsets
by their performance under different environmental conditions.
An illustrative example highlighting this feature is shown in
\figurename~\ref{fig:fig_2D_path}.
%}

\begin{comment}
\add{
After detecting a path, we look within the path to 
identify interesting subpopulations. Subpopulation in form of a path helps us to 
understand about the data which is hard to understand using simple scatter plot. 
For instance, \figurename~\ref{fig:fig_2D_path} shows scatter plot of points 
with respect to DAP (Days After Planting) and humidify where each point is a 
plant growth rate data cultivated in two locations (Nebraska and Kansas) in U.S. 
Each point is colored based on its location which indicates a clear separation 
of data based on humidity value between two locations. This coarse-level 
correlation is easy to test using traditional correlation test that shows a 
global trends. On the other hand, our method visualize the object as a graph and 
capable to detect subpopulations (in form of a path) from the graph. It 
helps us to understand the insight variability of the dataset. We 
overlaid the paths on the scatter plot (shown as arc in 
\figurename~\ref{fig:fig_2D_path}).
}
\end{comment}

%\add{
%In the paper,
We first define these features formally, and then present
algorithms to extract them from the topological objects constructed.  For
ranking purposes, we define a notion of interestingness.  
%}

\subsubsection{Software}
\label{sw} 
We have implemented our approach as a software tool, which we 
call \texttt{Hyppo-X} (stands for: \underline{Hyp}othesis extraction for \underline{p}hen\underline{o}mics). 
The tool is available as open source in the GitHub repository~\cite{hyppox}. 

Even though we demonstrate its utility in the context of
plant phenomics, our approach can be applied more broadly to other
similar application contexts where the goal is to identify interesting
subpopulations in general in an unsupervised manner from 
complex, high-dimensional biological data sets.

\section{Related work}\label{sec:related}
We are not aware of any other automated or semi-automated hypothesis extraction approaches for high-dimensional data sets. 
%Although presented in the context of plant phenomics data analysis, the application of our \texttt{Hyppo-X} framework can be extended to other domains which also have high-dimensional point cloud data sets. 
In what follows, we present some related work, both in topology and in plant phenomics, in order to put our contributions in context.

\subsection{Topology and Applications} 
There are several important properties that make algebraic topology
particularly effective for gleaning structural features out of
high-dimensional data.
First, topology studies shapes in a {\em coordinate-free way}, which
enables comparison among data sets from diverse sources or
coordinate systems.  
Second, topological constructions are {\em not sensitive to small
changes in data}, and robust against noise.
Third, topology works with {\em compressed representations} of
spaces in the form of \emph{simplicial complexes} (or 
triangulations)~\cite{Munkres1984}, which preserve information
relevant to how points are connected.  
Compared to more traditional techniques such as principal component
analysis, multidimensional scaling, manifold learning, and cluster
analysis, topological methods are known to be more sensitive to both
large and small scale patterns~\cite{Lumetal2013}.  

Topological data analysis (TDA) has been applied to a wide range of application
domains, in particular for visualization purposes~\cite{SiGh2007a,persistbarcodes04,PPintfaceBER06,geomsolvedelskoehl05,lukesubgrmine05,NiLeCa2011}.
The foundational work in TDA most relevant to this 
paper was done by Carlsson and coworkers~\cite{Lumetal2013}.
In~\cite{SiMeCa2007}, they describe a framework called \emph{Mapper} to
model and visualize high-dimensional data.
Most of this work has been on the visualization front.
A topology-based approach was also rated as the best overall entry at an
expression QTL (eQTL) visualization competition organized by the BioVis
community~\cite{bartlett_eqtl_2012}.

\subsection{Tools for Plant Phenomics}
Tools to decode associations between genotypes and phenotypes have been under development for over two decades.
These tools look at the genetic variation observed at one or more loci across the genome and study their correlation to quantitative traits.  
The techniques used can be summarized as follows: 
i) Linkage mapping usually begins with prior knowledge of the order of genetic markers and the goal of the mapping is to identify which markers co-segregate with a phenotype in a segregating population. It is usually used for traits controlled by fewer genes; 
ii) Quantitative Trait Locus (QTL) mapping that extends linkage to an interval of co-located markers along the genome; and 
iii) GWAS is typically used for traits controlled by many genes.  Typically all individuals within a diversity panel are scored for both genotypes at many markers AND phenotypes.  Statistical approaches are then used to identify statistically significant associations between markers and variation in trait values.
In relation to capturing environmental variability, efforts have been sparse.
\cite{brown2014traitcapture} presented an experimental framework supplemented by GWAS to model environmental effects on phenotypes.
\cite{lou2007generalized} provided a generalized linear model-based method to capture gene to environment interactions.  
In another related work, Yang {\it et al.} \cite{yang2017phenocurve} study the effect of environmental variables on photosynthesis efficiency in plants using a curve fitting approach. 

The approach presented in \emph{this} paper complements the above body of works in several ways including a new way to formulate the problem as one of unsupervised structure discovery, and in its method and capabilities (e.g., compact representation, visualization, and exploratory data analysis). 

\section{Hyppo-X: Our implementation of the Mapper framework}
\label{sec:Mapper}

\begin{figure}[!t]
\centering
\includegraphics[keepaspectratio=yes, width=\columnwidth]{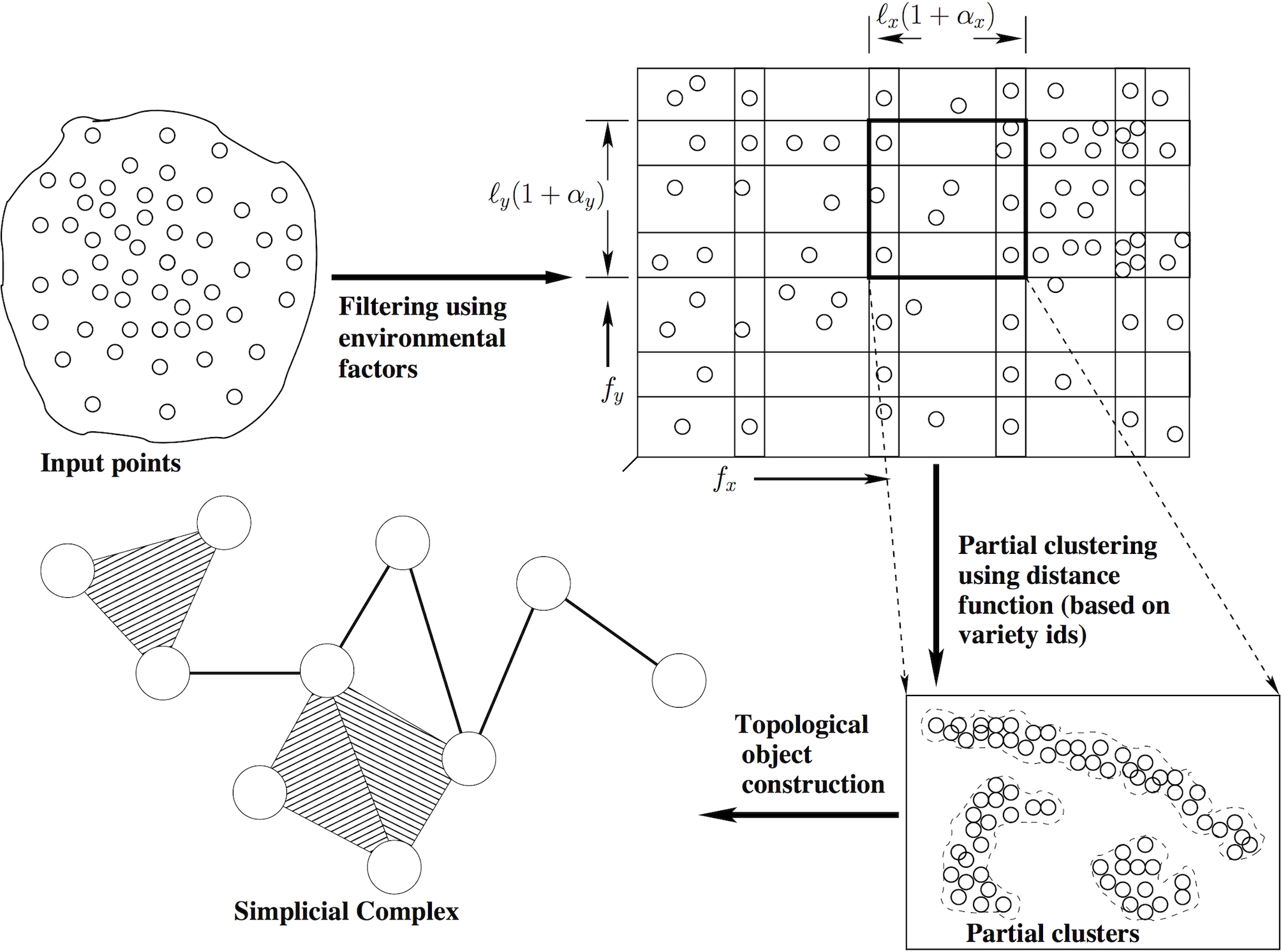}
\caption{\texttt{Hyppo-X} framework for analyzing phenomic data.}
\label{fig:Algo}
\end{figure}

The first step in our approach
is to construct topological representations using the connectivity properties of the data.
The motivation is to obtain higher order structural information about the
high-dimensional data prior to gleaning hypotheses. 
We present an implementation for the abstract Mapper algorithmic framework 
\cite{SiMeCa2007} for this purpose. 
In what follows, we describe the details of our implementations.
\figurename~\ref{fig:Algo} is a schematic illustration of our approach. 

\textbf{Input:} 
We are given a set of $n$ points $\mathcal{S}$ in a $d$-dimensional space, 
representing the space of interest $X$.
In the case of phenomics, a \emph{point} $x\in\mathcal{S}$ represents a crop individual that is measured at a particular time $t$, and the dimensions represent the attributes which describe the point at that time.
These include a set $E$ of $m$ factors  (e.g., time, temperature, humidity, etc.), and a performance trait, the phenotype $p$ (e.g., plant height or growth rate).
Note that these dimensions represent continuous variables 
(A point may also have other non-continuous or static variables (e.g., the genotype).
For the purpose of our topological representations we will use only the continuous
variables).

\textbf{Output:} 
We aim to create a highly compact coordinate-free 
representation of $X$ as a \emph{simplicial complex}, 
using a clustering (overlapping) of the points in $X$ (represented by $P$ here). 

\textbf{Simplicial complex:}
A \emph{simplicial complex} is a collection of
simplices (nodes, edges, triangles, tetrahedra, etc.) that fit together
nicely---all subsimplices of each simplex are included in the collection,
and any two simplices that intersect do so in a lower dimensional
subsimplex.  Specifically, each cluster is represented by a node
($0$-simplex).  Whenever two clusters have a non-empty intersection, we add an
edge ($1$-simplex), and when three clusters intersect, we add a triangle
($2$-simplex), and so on.

\smallskip
We now provide the main algorithmic details of the approach.

\begin{figure}[!t]
\centering
\includegraphics[keepaspectratio=yes, width=\columnwidth]{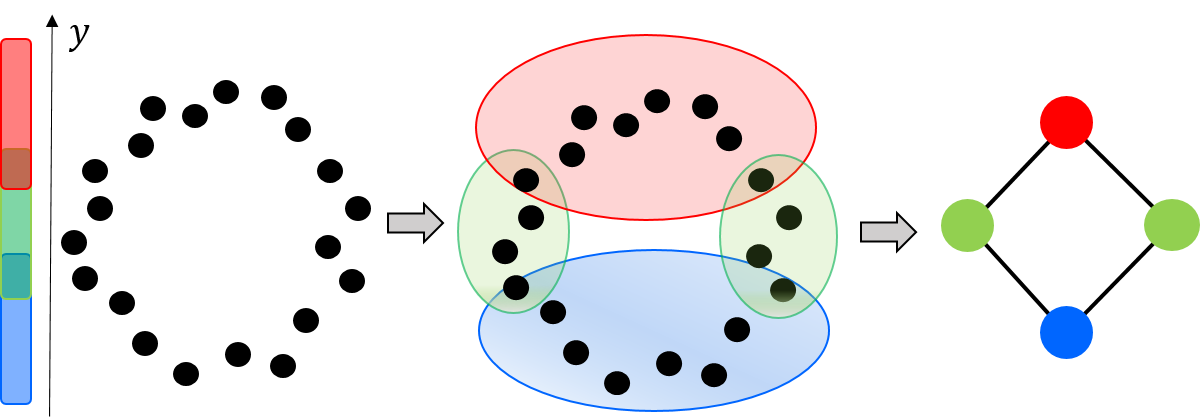}
\caption{The Mapper algorithm applied to a set of points sampled from a noisy circle.
We use the height of the points ($y$-coordinate) as the filter function.
We consider a cover of $Z \approx [-1,1]$ using $r=3$ overlapping intervals, with adjacent intervals overlapping roughly by a third (i.e., $g=33\%$).
The final Mapper is shown on the right.}
\label{fig:mapperillust}
\end{figure}

\subsection{Filtering}
\label{sec:Filtering}

The first component of the framework is a continuous function $f : X \to Z$ to a
real-valued parameter space $Z$, called the \emph{filter function}. 
For each factor $Z_j$, we define a filter function $f_j : X \to
Z_j$. We generate the open cover ${\cal U}_j = \{U_{ij}\}$ of $Z_j$ as follows:
%\begin{enumerate}

  $1$) {We divide each factor $Z_j$ into $n_j$ intervals
(``sub-regions''), each of length $\ell_j$. Thus the entire $d$-dimensional
region is divided into $n_1\times n_2\times\dots \times n_m$ subregions,
where each subregion represents a hyper-rectangle of area
$\ell_1\times\ell_2\times\dots\times\ell_m$.  Let the center of $i^{th}$
hyper-rectangle be $\{C_{1i},C_{2i},\dots,C_{mi}\}$.}  

  $2$) {We fix the center of each hyper-rectangle, and increase the
    length along each factor $Z_j$ by a certain
    percentage $\alpha_j$ such that an overlapping region is created
    between consecutive pairs of the open sets $U_{ij}$ and
    $U_{i+1,j}$, i.e., $U_{ij}\cap U_{i+1,j} \neq \emptyset$. After
    increasing the length of all sides in this fashion, the new area
    of the hyper-rectangle is $\ell_1(1+\alpha_1) \times \dots \times
    \ell_m(1+\alpha_m)$ 
	(See Section~\ref{sec:homology} for further explanation of how we choose the $\alpha_i$ 
	values).
    A 2D example is shown in \figurename~\ref{fig:Algo}.}
%\end{enumerate}

We formulate the efficient determination of individual point sets belonging to
each hyper-rectangle as a problem of range querying. Specifically, we implement
the following querying function:

\emph{{\bf Range Query:} Given $X$ and a hyper-rectangle $h$, return the subset
of points in $X$ that lie in $h$.}

To run this query efficiently, we use $k$-dimensional hyper-octtrees
\cite{aluru1999dynamic,clarkson1983fast}, which is a well known spatial data
structure that uses recursive bisection to index a spatially distributed set
of points. 
The compressed version of an $n$-leaves hyper-octree can be constructed in
$O(n\log n)$ time \cite{aluru1999dynamic}. Once constructed, a balanced binary
search tree that uses the order of the leaves is constructed. Using this
auxiliary data structure, in combination with the hyper-octree, enables an
$O(\log n)$ worst case search time for both point and cell searches
\cite{aluru1999dynamic}. To answer the regional query for a hyper-rectangle $h$,
we perform a top-down traversal of the hyper-octree by selectively retaining
only those paths that can include at least one point within $h$. This can be
achieved by keeping track of the corners of the cell defined by each internal
node in the tree. This approach ensures that each such query can be answered in
time that is bounded by the number of points in the hyper-rectangle.

\subsection{Generation of Partial Clusters}
\label{sec:Clustering}

Each open set (hyper-rectangle) computed by applying the 
filter functions is processed independently for generation of partial
clusters. The goal of clustering is to partition the set of points in each
hyper-rectangle based on their phenotypic performance.

Let $U$ represent an open set of points $\{x_1,x_2,\ldots x_t\}$.  Note that
each point $x\in U$ has a phenotypic trait value denoted by $p(x)$.  We
define a \emph{distance function} $d$ based on the phenotypic values of
points in $U$ as follows.  Given two points with trait values $p(x_i)$ and
$p(x_j)$, the distance $d(i,j)=|p(x_i)-p(x_j)|$.  

Given $U$ and distance function $d$, a \emph{partial clustering} is defined
by a partitioning of the points in $U$. 
We denote the set of partial clusters resulting from any given open set $U$
as $\mathcal{C}_{U}$.
Subsequently, we denote the set of all partial clusters formed from \emph{all} open sets (hyper-rectangles) by 
$\mathcal{C}=\bigcup\limits_{U} \mathcal{C}_U$.

For the purpose of clustering, any distance-based clustering method can be applied.
We implemented a density-based clustering approach very similar to that of DBSCAN \cite{ester1996density}.
It covers two key points: 
a) the set of partial clusters generated from within a hyper-rectangle
represents a partitioning of those points; and 
b) two partial clusters generated from a pair of adjacent (overlapping)
hyper-rectangles could potentially have a non-empty intersection in points.
In fact it is this intersection that renders connectivity among the partial
clusters generated, the information for which will be used in the subsequent
step of simplicial complex generation.

\subsection{Construction of Simplicial Complexes}
\label{sec:Construction}

From the set of partial clusters $\mathcal{C}$, we construct a simplicial complex $M$ as follows.
We describe the details for the 2D case, where no more than four open
sets (hyper-rectangles) can mutually intersect. The extension to
higher dimensions is straightforward.
Starting with an empty simplicial complex, we implement the following steps:
%\begin{enumerate}

  $1$) {A $0$-simplex (or vertex) is added to the simplicial complex $M$ for every partial 
  cluster.}
  
  $2$) {Next, a $1$-simplex (edge) is added to $M$ for every non-empty $2$-way
intersection between any two partial clusters. Note that such intersections
could exist only between partial clusters originating from different open
sets.}

  $3$) {Following the same procedure as above, we also add $2$-simplices
(triangles) and $3$-simplices (tetrahedra) to $M$ by enumerating only those
$3$-way and $4$-way intersections, respectively, that could be non-empty.} 
%\end{enumerate}

The required multi-way intersections are computed using the range querying
function described earlier (in Section~\ref{sec:Filtering}).

The Mapper algorithm \cite{SiMeCa2007} produces highly compressed visual representations of high-dimensional data that reveal significant structural aspects.
For example, consider the instance where $X$ is a set of points in $\R^2$ sampled from a noisy unit circle (see \figurename~\ref{fig:mapperillust}).
We use the height of the points (i.e., their $y$-coordinate values) as the filter function.
We consider a cover of $Z$, which is almost $[-1,1]$, into $r=3$ overlapping intervals, with adjacent intervals overlapping roughly by a third (i.e., $g=33\%$).
%The interval values increase from low to high as the color changes from blue to red.
The pullback cover of $X$ then has four pieces, with the subset of points with height in the middle interval forming two connected components.
We then use the Euclidean distance between the points (in $\R^2$) as the distance function to cluster the points in each component using, e.g., single linkage clustering.
Thus we get one node per component, which we color from blue to red according to the mean height of the points in each node.
We also get four connecting edges capturing the overlap of the clusters.
Note that we go from around $20$ points in $X$ to just four nodes and four edges in the Mapper.
At the same time, this highly compact representation captures the underlying structure of $X$---the circle.

\subsection{Graph Formulation} \label{sec:graph_form}

We construct a weighted directed graph $G=(V,E)$ representation of the $1$-skeleton of $M$ along with some additional information.
We set $V$ as the set of vertices ($0$-simplices) of $M$, and $E$ as the set of edges ($1$-simplices) of $M$.
We assign directions and weights to the edges as follows.
Each vertex $u \in V$ denotes a subset of points from $X$ that constitute a partial cluster.
We denote this subset as $X(u)$. 
We let $g(u)$ and $f_i(u)$ denote the average values of the clustering function $g$ (dependent variable) and the filter function $f_i$, respectively, for all points in $u$:
\begin{equation}\label{eq:gu}
  g(u)=\frac{\Sigma_{x \in X(u)} \, {g(x)}}{|X(u)|}
\end{equation}
and
\begin{equation}\label{eq:fu}
   f_i(u) = \frac{\Sigma_{x \in X(u)} \, {f_i(x)}}{|X(u)|}\,, ~i=1,\dots,h.
\end{equation}
For an edge $e=(u,v)$ in $E$, we assign as its weight as:
$\omega(e) = |g(u)-g(v)| \,$.
%For an edge $e=(u,v)$ in $E$, we assign as its weight the absolute difference between the average cluster function values of the two vertices:
%$\omega(e) = |g(u)-g(v)| \,$.
Notice $\omega(e) \geq 0$ for all edges $e$ in $G$.
In addition, the direction of the edge $e$ is set from the lower weight vertex to the higher weight vertex---i.e., if $\omega(u)\leq \omega(v)$ then $e:u \to v$, and $e:v \to u$ otherwise.
We let $n=|V|$ and $m=|E|$ denote the numbers of vertices and edges in $G$, respectively.

\begin{figure}[!t]
\centering
\includegraphics[keepaspectratio=yes, width=2.5in]{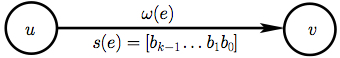}
\caption{
An edge $e$ between two intersecting partial clusters (nodes $u$ and $v$).
The direction of the edge indicates the direction in which the mean
phenotypic/performance value increases. 
The signature $s(e)$ is a $k$-bit vector that captures the directions of 
change for each of the $k$ filter functions (e.g., environmental variables) 
along the edge---0 implies decreasing and 1 implies increasing. 
The $i^{th}$ bit corresponds to the $i^{th}$ filter function.}
\label{fig:Edge}
\end{figure}

\subsubsection{Edge}\label{sec:edge}

If the simplicial complex was constructed using $h$ out of the
$m$ continuous variables (as filter functions), then along each edge,
each continuous variable $f_i$ can independently increase or
decrease.  Since we are trying to link the change of each of these variables
relative to the change in phenotype (along an edge), we record a $h$-bit
\emph{signature} for each edge.  

We assign a $h$-bit binary signature $\Sig(e)=b_1b_2 \dots b_h$ to the oriented edge $e=(u,v)$ (i.e., $e: u \to v$) to capture the covariation of $g$ and the filter functions $f_i$.
We set $b_i=1$ if $f_i(u) \leq f_i(v)$, and $b_i=0$ otherwise.

In other words, let an edge's direction be $u\rightarrow v$. Then, if the
mean value for the continuous variable $f_i$ increases from $u$ to $v$,
then the corresponding signature bit is $1$; and $0$ otherwise.

\figurename~\ref{fig:Edge} illustrates a directed, signed edge in our representation.

Note that based on the above edge definition, there cannot be any cycles in $G(V,E)$, making it a Directed Acyclic Graph (DAG).

\subsection{Persistent homology}
\label{sec:homology}
We employ the concept of persistent homology~\cite{EdLeZo2002} to choose the
final topological object for further analysis. In particular, the method in
which overlapping intervals are chosen (by specifying growing overlap
percentages $\alpha_i$, see Section~\ref{sec:Filtering}) is already guided by
this principle. Termed \emph{multiscale mapper}, growing the intervals in this
fashion ensures the topological objects formed (at each set of growing
$\alpha_i$ values) satisfy a monotonic inclusion property~\cite{DeMeWa2016}.
Hence results from persistent homology could be used to guarantee (theoretical)
stability of the topological object formed (in the sense of persistence). At the
same time, no implementation of multiscale mapper is known. Instead, we increase
each $\alpha_i$ in steps of $2.5\%$, and construct the topological objects for
each set of $\alpha_i$ values. We then construct the persistence barcodes (in
dimensions $0,1$, and $2$) using the sequence of topological objects formed by
employing JavaPlex, a standard software tool for this purpose.
We then pick $\alpha_i$'s such that all three barcodes do not change for values
at or higher than the chosen cutoff, ensuring the corresponding topological
object chosen is indeed stable. 

\section{Extracting interesting features}
\label{sec:IntFeature}

%\add{
In this section, we define two features---flares and paths---that can be extracted from
the topological representations we construct (in Section~\ref{sec:Mapper}), and subsequently describe our algorithms to extract those features. 
Interesting flares and paths hold two different types of information: 
flares are more useful to identify subpopulation divergence, whereas paths are more useful to identify and analyze subpopulations over the performance spectrum.
%}

\subsection{Interesting Flares}\label{sec:IF}
We propose a framework to detect and use ``flares'' (defined below) that characterize branching phenomena in phenomics data sets.
  
%\subsubsection{Method}\label{sec:IFMethod}
We first construct a directed graph $G$ based on the process discussed in Section~\ref{sec:graph_form}.
Given an edge $e=\{u,v\}$, we direct the edge by default from the cluster showing a lower phenotypic performance as measured by its mean phenotypic value to one with the higher mean phenotypic value (see \figurename~\ref{fig:Edge}).
This scheme allows us to track a trail of clusters that show an improving trajectory in performance by a user-selected phenotypic trait (e.g., yield or plant height).
In order to capture branching phenomena effectively, we modify this directing procedure by using mean phenotypic values of \emph{subset of individuals belonging to shared genotypes} between nodes $u$ and $v$.
%---see \figurename~\ref{fig:Edge}.
%and text following the same, as well as the illustration in Figure \ref{fig:1D_Edge_Dir}).

%{\bf Flares:}

\begin{definition} \label{def:stbranch}
  A \emph{source} (\emph{terminal}) node in a directed graph is one that has no incoming (outgoing, respectively) edges. 
  A \emph{branching} node in a directed graph is one that has at least two outgoing edges.
\end{definition}

Note that, by the above definitions, a source node can also be potentially a branching node.
Furthermore, we use the term \emph{simple path} to refer to a path in the graph in which no node, with the possible exception of the sentinel nodes (beginning and ending) of the path, is a branching node.  

\begin{figure}[htp!]
\centering
\includegraphics[keepaspectratio=yes, width=0.5\textwidth]{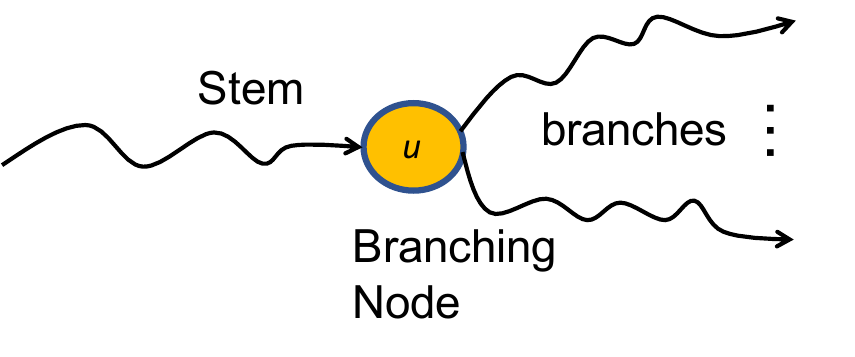}
\caption{An illustration of a flare.}
\label{fig:Flare}
\end{figure}

We define a \emph{stem} and a \emph{branch} associated with a branching node as follows (see Figure~\ref{fig:Flare} for an illustration).
\begin{definition} \label{def:stem}
  Given a branching node $u$, a \emph{stem} is a possibly empty simple path that ends in $u$.
\end{definition}

  Note that there can be multiple stems %that can be constructed for paths
  ending at a branching node $u$.
  There are two classes of such stems---those that are entirely non-overlapping (i.e., simple paths ending at $u$ that are otherwise node-disjoint) and those that are nested (i.e., they originate from different starting nodes in the same parent simple path ending at $u$). 

\begin{definition} \label{def:branch}
  Given a branching node $u$, a \emph{branch} refers to a non-empty path (simple or not) that originates at $u$.
\end{definition}
  Note that two branches originating at the same branching node can possibly intersect. 
  Furthermore, there are at least two branches originating at a branching node (by definition of a branching node).

Let $B(u)$ denote the set of all branches originating at a branching node $u$ and
$S(u)$ denote the set of non-overlapping (i.e., non-nested) stems ending at $u$.

\begin{definition} \label{def:flare}
We define a \emph{flare} to be a unique combination of a branching node $u$, a stem $s\in S(u)$, and a subset $B^\prime(u) \subseteq B(u)$.
Here, we do \emph{not} enforce that a stem be non-empty, to allow detection of flares strictly originating at a given branching node.
However, we \emph{do} enforce that each branch selected is non-empty (i.e., has at least one edge) and that the subset selected $B^\prime(u)\subseteq B(u)$ contains at least two or more branches (as illustrated in \figurename~\ref{fig:Flare}).
\end{definition}

\begin{figure}[htp!]
\centering
\includegraphics[keepaspectratio=yes, width=0.6\textwidth]{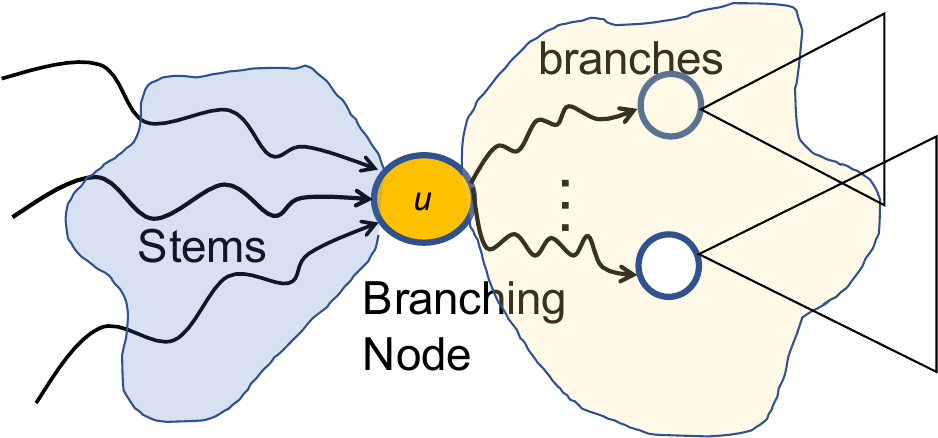}
\caption{Conceptual illustration of how flares are constructed from a given branching node $u$.
	Stems are selected from the set of incoming simple paths, and branches are selected from the DAGs rooted at $u$. 
	The boundaries of the selection are determined by ``casting a net'' on either side of $u$ and including all ``areas'' where there is shared individual coverage.See text above for further details.}
\label{fig:FlareFishNet}
\end{figure}

The selection of the stem and branches to include in a flare is computed deterministically as a function of the branching node.
Intuitively, the idea is to examine the set of individuals ``covered'' by the branching node, 
and then ``cast a net'' in either direction, on all simple paths leading up to $u$ (candidate stems) and on all the branches originating at $u$, 
as far as there is a non-empty intersection with the individual set of the branching node
(see \figurename~\ref{fig:FlareFishNet}).

The rationale for this selection scheme is as follows. 
In an application such as phenomics, each ``point'' included in a cluster is typically a given plant crop (``individual'') observed in certain time and space.
Therefore, by the way we construct our topological object using intersections between adjacent clusters, the same individual may continue to appear in a sequence of clusters (i.e., in a path) on either side of a branching node.
Therefore, by considering the set of individuals covered by a branching node, and examining how that set distributes itself across the branches, we can discover interesting subpopulation-level variations (or differences in the way they respond to various environmental filters). 
In a population where there is also a large genetic diversity, one can adapt the same procedure to include the set of genotypes covered (instead of plant individuals).

{\bf Detection of flares:}
More formally, let $N(u)$ denote the set of individuals covered in the cluster
corresponding to $u$. Then, we follow the trail of clusters in either direction to
incrementally grow the corresponding stem or branch, as follows. 
For stem computation, we enumerate all the simple paths ending at $u$, and for each such
simple path (candidate stem), we begin at the node $v$ which is the immediate predecessor
of $u$ and compute $N(v)\cap N(u)$. If the intersection is non-empty then we include $v$ in
the current stem and iteratively walk to the next predecessor (until either the simple path
terminates or the intersection becomes empty). Note that at each step, we compute the
intersection with $N(u)$.  

A similar procedure is carried out to enumerate all branches originating at $u$, walking
forward instead of backward, with the caveat that we do not need to restrict the elongation
process to only simple paths in the forward direction. In other words, if we encounter
another branching node, the algorithm proceeds recursively,
except that at every subsequent step going forward from the second branching node,
the intersection is computed only relative to the original branching node $u$.  

Note that the above procedure is deterministic, in that given a branching node, the reach
of a flare involving that branching node is determined by the reach of the set of individuals in $u$ on either side of $u$ in the DAG. 
In fact, this procedure would also detect \emph{all} the flares involving $u$. 
%Let $S(u)$ denote all the non-overlapping stems detected and $B^\prime(u)$ denote all the distinct
%(but possibly overlapping) branches of $u$. Then,
More precisely, the cross-product of $S(u)$ and $B^\prime(u)$ (as specified in Definition~\ref{def:flare}) yields the set of all flares involving $u$.

%Note that every flare will represent a connected subgraph pivoted on a branching node (as shown in \figurename~\ref{fig:Flare}).

{\bf Scoring flares:}
In order to compare and relatively rank flares, we devise a simple scheme to score each flare.
Given a flare $f$, we compute its ``interestingness score'' as follows. 

First, we associate a weight to all edges. The weight of an edge is given by the absolute difference in the phenotypic performance (cluster means) between the two corresponding clusters.
Intuitively, the larger the performance variation, the more interesting that edge is to a branch.
Note that since we use the absolute value of the difference, all edge weights are positive.

\begin{figure}[!t]
  \centering
  %\vspace*{-0.01in}
  %\includegraphics[keepaspectratio=yes, width=2.5in]{figGatherScatter.pdf}
  \includegraphics[keepaspectratio=yes, width=0.6\textwidth]{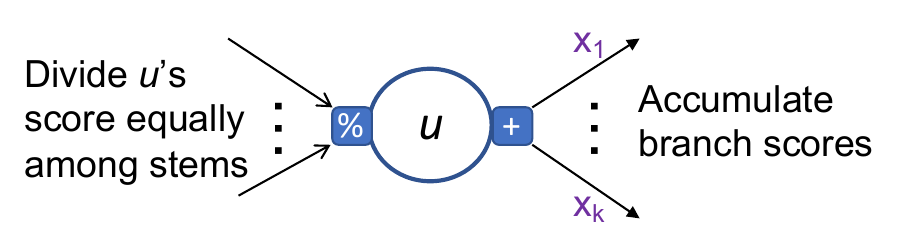}
  %\vspace*{-0.1in}
\caption{
Illustration of how the interestingness score propagates through a flare.
Computation proceeds as an accumulation process from the branches to the stems. 
}
\label{fig:GatherScatter}
\end{figure}

We score the flare using its edge weights as follows (see \figurename~\ref{fig:GatherScatter}).
Note that there is a unique subgraph induced by each flare and that subgraph also will be acyclic (as it is derived from a DAG).
Therefore, we perform a simple bottom-up/post-order traversal of that induced DAG, starting
at each terminal node and climbing up the parent and the ancestor levels. 
At each step, we perform a simple gather-scatter way to propagate the scores across levels.
More specifically, at a node $u$, all the scores of its child branches are added
(``gather''), and the value is then equally divided (``scatter'') among its predecessor
branches. The algorithm terminates when it reaches the main branching node $u$ of this
flare. 
Once scored, the flares can be rank ordered in the decreasing order of score and displayed.

In our current implementation, only branches contribute to the score of a flare at a
branching node. Stems do not contribute, the rationale being that examining the branches
typically suffices for explaining how a population, covered at the branching node, diverges. 
However, the procedure can be extended to include stem scores as needed. 
The information contained in the stem is still useful during our subsequent analysis and
interpretation.

  Our procedure for scoring flares takes running time linear in the size of the flare.

\subsection{Interesting Paths}\label{sec:IP}
We propose a framework to detect and use ``paths'' (defined below) that helps to identify interesting subpopulation in phenomics data sets.

\begin{definition} \label{def:IPath}
An \emph{interesting $k$-path} for a given $k$ with $1 \leq k \leq n-1$ is a directed path $P=[e_{i_1}, \dots, e_{i_k}]$ of $k$ edges in $G$, such that $\Sig(e_r)$ is identical for all $r=i_1,\dots,i_k$.
  An \emph{interesting path} is a path of arbitrary length in the interval $[1,n-1]$.
\end{definition}  

\begin{definition} \label{def:IScr}
  Given an interesting $k$-path $P=[e_{i_1}, \dots, e_{i_k}]$ in $G$ as specified in Definition \ref{def:IPath}, we define its \emph{interestingness score} as follows.
  \begin{equation} \label{eq:IScr}
    \IScr(P) = \sum_{r=1}^k \omega(e_{i_r}) \times \log(1+r) \, 
  \end{equation}
  In particular, the contribution of an edge $e \in P$ to $\IScr(P)$ is set to $\omega(e) \times \log(1+\rank(e,P))$, where $\rank(e,P)$ is the rank or order of edge $e$ as it appears in $P$.
\end{definition}

%\begin{remark}
Intuitively, we use the rank of an edge as an inflation factor for its weight---the later an edge appears in the path, the more its weight will count toward the interestingness of the path.
This logic incentivizes the growth of long paths.
The log function, on the other hand, helps temper this growth in terms of number of edges.
%One could use, for instance, the rank of the edge in place of the log function.
%\end{remark}

{\bf Optimization Problems: }
We now present multiple optimization problems with the broader goal of identifying interesting path(s) that maximize interestingness score(s).

{\bfseries \maxip}:~Find an interesting path $P$ in $G$ such that $\IScr(P)$ is maximized.

{\bfseries \ip}:~Find a collection $\cP$ of interesting paths in $G$ such that the total interestingness score $\IScr(\cP) = \sum_{P \in \cP} \IScr(P)$ is maximized ($\cP$ will exactly cover $E$, i.e., each $e \in E$ is part of exactly one $P \in \cP$).

%{\bfseries \kip}:~For a given $k$ between $1$ and $n-1$, find a collection $\cP$ of interesting $k$-paths such that each $e \in E$ is part of at most one $P \in \cP$, and the total interestingness score $\IScr(\cP) = \sum_{P \in \cP} \IScr(P)$ is maximized.

\smallskip
A detailed analysis the above optimization problems (and related variants) with their respective complexity results and proofs are provided in a separate manuscript~\cite{KaKaKr2017}.
  %kalyanaraman2017interesting}. 
	In what follows, we present an exact algorithm for the \maxip~problem.
	We also present an efficient heuristic for the \ip~problem.
	Both these algorithms are implemented in our \hyppox{} framework. 
%}

\subsubsection{The \maxipsp Problem} \label{sec:MaxIP}

The goal of \maxipsp is to identify an interesting path with the maximum interestingness score. 
We show \maxipsp is P on directed acyclic graphs (DAGs).

%\paragraph{\maxipsp on directed acyclic graphs}  \label{ssec:MaxIPDAG}

\begin{lemma}\label{lem:MAXIP_DAG_in_P}
 \maxipsp on a directed acyclic graph $G=(V,E)$ is in P. 
\end{lemma}

\begin{proof}
  We present a polynomial time algorithm for \maxipsp on a DAG (as proof of Lemma~\ref{lem:MAXIP_DAG_in_P}).
  The input is a DAG $G=(V,E)$ with $n$ vertices and $m$ edges, with edge weights $\omega(e)\geq 0$ and signatures $\Sig(e)$ for all $e\in E$.
  The output is an interesting path $P^*$ which has the maximum interestingness score in $G$. 
  We use dynamic programming, with the forward phase computing $\IScr(P^*)$ and the backtracking procedure reconstructing a corresponding $P^*$.

%\delete{Delete this tabular part\\START}
% NO, WE NEED THIS. 
  Let $T(i,j)$ denote the score of a maximum interesting path of length $j$ edges ending at edge $e_i$ for $i\in [1,m]$. 
  Since an interesting path could be of length at most $(n-1)$, we have $j\in [1,n-1]$.
  Therefore the values in the recurrence can be maintained in a 2-dimensional table of size
  $m \times (n-1)$, as illustrated in \figurename~\ref{fig:maxipalgoTbl}. 
  %, where cell $T[i][j]$ stores the value for $T(i,j)$.
The algorithm has three steps:
\begin{itemize}
\item{\bf Initialization:}
$ T(i,1) = \omega(e_i) \times \log(2)~, \mbox{where } 1\leq i\leq m$.
%We initialize the first column of the table $T$ as follows.
%\[ T(i,1) = \omega(e_i) \times \log(2)~(1\leq i\leq m).\]

\item{\bf Recurrence:}
For an edge $e=(u,v) \in E$, we define a \emph{predecessor edge} of $e$ as any edge
$e^\prime\in E$ of the form $e^\prime=(w,u)$ and $\Sig(e^\prime)=\Sig(e)$. 
Let $\Pred(e)$ denote the set of all predecessor edges of $e$.
Note that $\Pred(e)$ can be possibly empty.
We define the recurrence for $T(i,j)$ as follows.
\begin{equation} \label{eq:rec}
  T(i,j) = \max_{e_{i^\prime} \in \,\Pred(e_i)}\big\{T(i^\prime,j-1) + \omega(e_i) \times \log(1+j) \big\}
  %T(i,j) = \max_{e_{i'} \in \,\Pred(e_i)}\big\{T(i',j-1) + \omega(e_i) \times \log(1+j) \big\},~j=2,\dots,n-1.
\end{equation}

\item{\bf Output:}
  We report the score that is maximum in the entire table. 
  A corresponding optimal path $P^*$ can be obtained by backtracking from that cell to the  first column.

  %Let $(i_{\max},j_{\max}) = \arg\max_{i,j}\{T(i,j)\}$.
  %Then the maximum interestingness score for \maxipsp is $T^* = T(i_{\max},j_{\max})$.
  %We obtain an optimal path $P^*$ with $\IScr(P^*)=T^*$ by backtracking from $T(i_{\max},j_{\max})$.

\end{itemize}
%\delete{END}

	\subparagraph{\underline{Proof of correctness:}}
%\begin{proofofcorrectness}
Any interesting path in $G$ can be at most $n-1$ edges long.
As a particular edge could appear anywhere along such a path, its rank can range between $1$
and $n-1$.
Hence the $m \times (n-1)$ recurrence table $T$ sufficiently captures all possibilities for each edge in $E$.
%We make the following observation about the structure of maximum interesting paths identified by the dynamic programming algorithm, which guarantees its correctness.
The following key observation completes the proof. 
Let $P^*(i,j)$ denote an optimal scoring path, if one exists, of length $j\in [1,n-1]$ ending at edge $e_i\in E$.
If $P^*(i,j)$ exists and if $j>1$, then there should also exist $P^*(i^\prime,j-1)$ where $i^\prime\in \Pred(e_i)$.
Furthermore, the edge $e_i$ \emph{could not} have appeared in $P^*(i^\prime,j-1)$ because $G$ is acyclic. 
Therefore, due to the edge-disjoint nature of $P^*(i^\prime,j-1)$ and the remainder of $P^*(i,j)$ (which is $e_i$), the principle of optimality is preserved---i.e., the maximum operator in Eqn.~(\ref{eq:rec}) is guaranteed to ensure optimality of $T(i,j)$.
%\end{proofofcorrectness}
  
\begin{figure}
  \centering
  \includegraphics[keepaspectratio=yes, width=0.8\columnwidth]{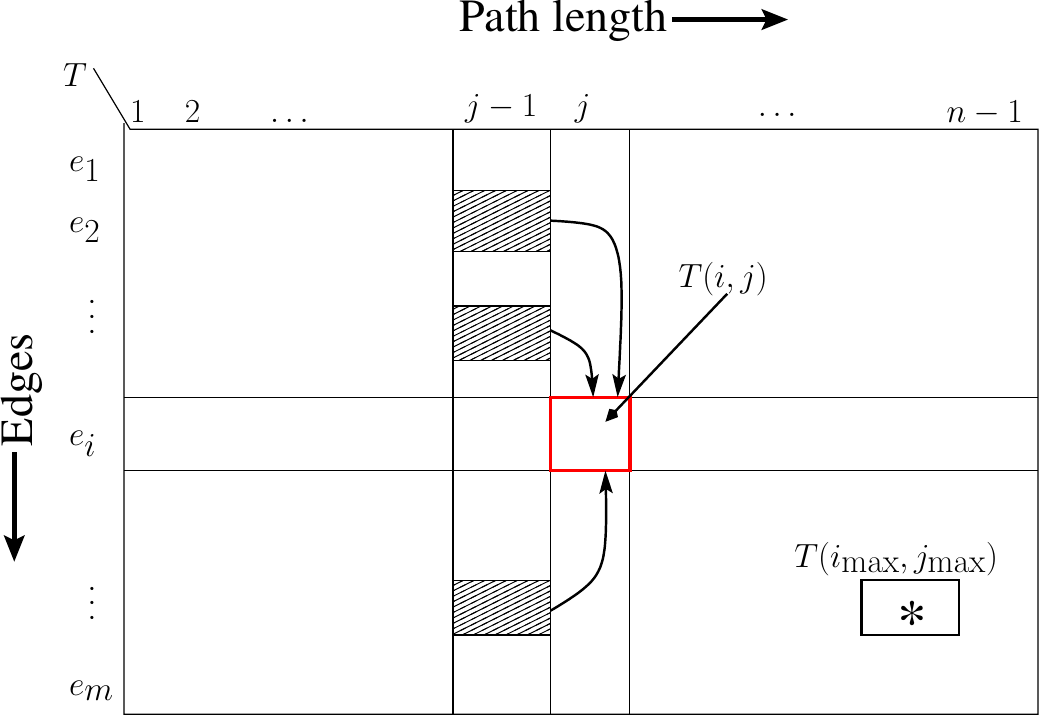}
  \caption{Table $T(i,j)$ for the \maxipsp algorithm.} 
  \label{fig:maxipalgoTbl}
\end{figure}

%%%%%%%%%% End here %%%%%%%%%%%%%%%

%\vspace*{-0.2in}
\subparagraph{\underline{Complexity analysis:}}
%\delete{Modify this part based on efficient algo}
The above dynamic programming algorithm can be implemented to run in $O(mn)$ space and a worst-case time complexity of $O(mnd_{\rm in})$, where $d_{\rm in}$ denotes the maximum indegree of any vertex in $V$. 
%We can initialize the table $T$ containing $m$ rows and $n-1$ columns, and compute the values in the table one column at a time, starting from the first column to the last column. 
%Since the maximum number of predecessors of an edge is bounded by $d_{\rm in}$, the cost of computing each cell in the table $T$ is $O(d_{\rm in})$.
%Therefore, the overall runtime complexity is $O(mnd_{\rm in})$.
%Since $d_{\rm in} < n$ in any directed graph, we get a worst-case time complexity of $O(mn^2)$, showing that \maxipsp on a DAG is in P.
%At the same time, $d_{\rm in}$ could be much smaller than $n$ in specific cases.
%In the case where $d_{\rm in}$ is a constant, the algorithm runs in $O(mn)$ time.
\end{proof}

\subparagraph{\underline{Algorithmic improvements:}} \label{sssec:algoimpr}
%\delete{Use this algorithm instead of DP using table}
The above dynamic programming algorithm for \maxipsp for DAGs can be implemented to run in space and time smaller in practice than the worst case limits suggested above. 
%We present the key ideas for such an implementation. 
First, we note that computing the full table $T$ is likely to be wasteful, as it is likely to be sparse in practice.
The sparsity of $T$ follows from the observation that an interesting path of length $j$ ending at edge $e_i$ can exist only if there exists at least one other interesting path of length $j-1$ ending at one of $e_i$'s predecessor edges. 
We can exploit this property by designing an iterative implementation as follows.

Instead of storing the entire table $T$, we store only the rows (edges), and introduce columns on a ``need basis'' by maintaining a dynamic list $L(e_i)$ of column indices for each edge $e_i$.
%\smallskip
\begin{compactenum}[S1)]
  \item
    Initially, we assign $L(e_i)=\{1\}$, as each edge is guaranteed to be in an interesting path of length at least $1$ (the path consisting of the edge by itself).  
  \item \label{stepupdate}
    In general, the algorithm performs multiple iterations; 
    within each iteration, we visit and update the dynamic lists for all edges in $E$ as follows. 
    For every edge $e_{i'}\in \Pred(e_i)$, $L(e_i)=L(e_i) \cup \{ \ell + 1 \, | \, \ell\in L(e_{i'})\}$.
    The algorithm iterates until there is no further change in the lists for any of the edges. 
\end{compactenum}

\smallskip
\noindent The number of iterations in the above implementation can be bounded by the length of the longest path in the DAG (i.e., the diameter $\delta_{\max}$), which is less than $n$. 
Also, we implement the list update from predecessors to successors such that each edge is visited only a constant number of times (despite the varying products of in- and out-degrees at different vertices). 
To this end, we implement the update in S\ref{stepupdate} as a two-step process:
first, performing a union of all lists from the predecessor edges of the form $(*,v)$ so that the merged lists can be used to update the lists of all the successor edges of the form $(v,*)$. 
Thus the work in each iteration is bounded by $O(m)$. 

Taken together, even in the worst-case scenario of $(\delta_{\max}+1)$ iterations, the overall time to construct these dynamic lists is $O(m\delta_{\max})$. 
Furthermore, during the list construction process, if one were to carefully store the predecessor locations using pointers, then the computation of  the $T(i,j)$ recurrence in each cell can be executed in time proportional to the number of \emph{non-empty} predecessor values in the table. 
Overall, this revised algorithm can be implemented to run in time $O(m \delta_{\max} d_{\rm in})$, 
and in space proportional to the number of non-zero values in the matrix. 

Further, the above implementation is also inherently parallel since the list value at an edge in the current iteration depends only on the list values of its predecessors from the previous iteration.
%Therefore, we can implement the algorithm in a parallel fashion, further enhancing its efficiency in practice.

%\begin{comment}

\subsubsection{An Efficient Heuristic for \ipsp} \label{sec:IPheuristic}
%\add{
In addition to an exact algorithm for \maxip{} (Section~\ref{sec:MaxIP}), we also present
an efficient heuristic for finding \ip. The \ip{} formulation aims at identifying a \emph{set} of
edge-disjoint interesting paths in $G$ such that the overall sum of their scores is maximized.
\ip{} is relevant in contexts where the user is interested not only in the maximum-scoring path but
also multiple others that cover different parts (and hence different subpopulations) of $G$. 
Once identified, these paths can be rank ordered in descending order of their scores for display purposes.
%}

%\add{
Algorithm~\ref{alg:GreedyIP} shows the pseudocode for our \ip{} heuristic. 
The approach is a simple greedy strategy, in which we iteratively find 
the next best scoring path (by calling \maxip), add it to the working set of paths, and remove all
edges of that path from the graph. This procedure is carried out until there are no more edges left. 
The algorithm has a worst-case runtime complexity of $O(m^2 \delta_{\max} d_{\rm in})$.
%}

 \begin{algorithm}[tbh]
      \KwIn{DAG $G=(V,E)$ with $\omega(e),\,\Sig(e) ~\forall e \in E$
      }
      \KwOut{A set of edge-disjoint interesting paths $\cP$ in $G$}
      $\mathcal{P} = \emptyset$\\
      \Repeat{$E=\emptyset$}
             {
               $P\gets $ Compute \maxipsp on $G=(V,E)$ and return a most interesting path\\
               $\mathcal{P}\gets \mathcal{P}\cup \{P\}$\\
               Remove edges in $P$ from $E$\\
             }
      \Return $\mathcal{P}$
      \caption{Greedy Heuristic for \ipsp on DAGs}
      \label{alg:GreedyIP}
    \end{algorithm}

%\end{comment}

\section{Experimental evaluation}\label{sec:result}
In our experiments, we used two real-world maize data sets. 
For the first batch of experiments described in Section~\ref{sec:data1}, we used a maize data set containing
growth information of two maize genotypes that were cultivated in two 
different locations in the U.S. (Kansas and Nebraska).
We refer to this data set as the ``KS/NE'' data set.
We used this data set to test various functionalities of our \hyppox{} framework 
%including the evaluation of flares (Section~\ref{sec:flare-result}) and interesting paths (Section~\ref{sec:path-results}). 
including hypotheses extraction in the case of single filter function (Section~\ref{sec:sfFltrFnc}) 
and two filter functions (Section~\ref{sec:dfFltrFnc}) using both flares and paths.
For the second batch of experiments, we used another maize data set collected 
from two field locations in Nebraska that had identical conditions except for one environmental parameter---one location was irrigated while the other was not.  
This data set covers individuals from $80$ genotypes (as described in Section~\ref{sec:irr-result}).
We refer to this data set as ``irrigation-controlled'' data set.

\subsection{KS/NE data set}\label{sec:data1}
%We used a real-world maize data set to test and evaluate the proposed algorithmic framework described in Section~\ref{sec:IntFeature}.
This maize data set consists of phenotypic and environmental measurements for two genotypes (abbreviated here for simplicity as $A$ and $B$), grown in two geographic locations (Nebraska (NE) and Kansas (KS)). 
The data consists of \emph{daily} measurements of the genotypes' growth rate alongside multiple environmental variables, over the course of the first $100$ days of the growing season.
For the purpose of our analysis we treat each unique [genotype, location, time] combination as a ``point''. 
Consequently, the above data set consists of $N=400$ points.
Here, ``time'' is measured in Days After Planting (DAP). 
An ``individual'' in this data set refers to a plant individual that corresponds
to a [genotype, location] combination.
Each point has one phenotypic value (observed growth rate) and $10$ environmental variables, including (among others) humidity, temperature, rainfall, solar radiation, soil moisture, and soil temperature.
%We studied a multitude of these environmental variables; the results presented in this paper correspond to humidity as it led to more interesting observations compared to the other variables.

To study flares and paths, 
we constructed topological objects out of the KS/NE data set, 
using single and two filter function(s).

\subsection{Single filter function}\label{sec:sfFltrFnc}
First, we constructed our topological object using DAP as the filter and used the difference in growth rates to calculate pairwise distances between points (in the clustering step).
This study is aimed at understanding how the population of individuals (of both genotypes in both locations) show varying trends in phenotypic performance (i.e., growth rate here) as a function of time. 

\begin{figure}[htp!]
\centering
\includegraphics[keepaspectratio=yes, width=\textwidth]{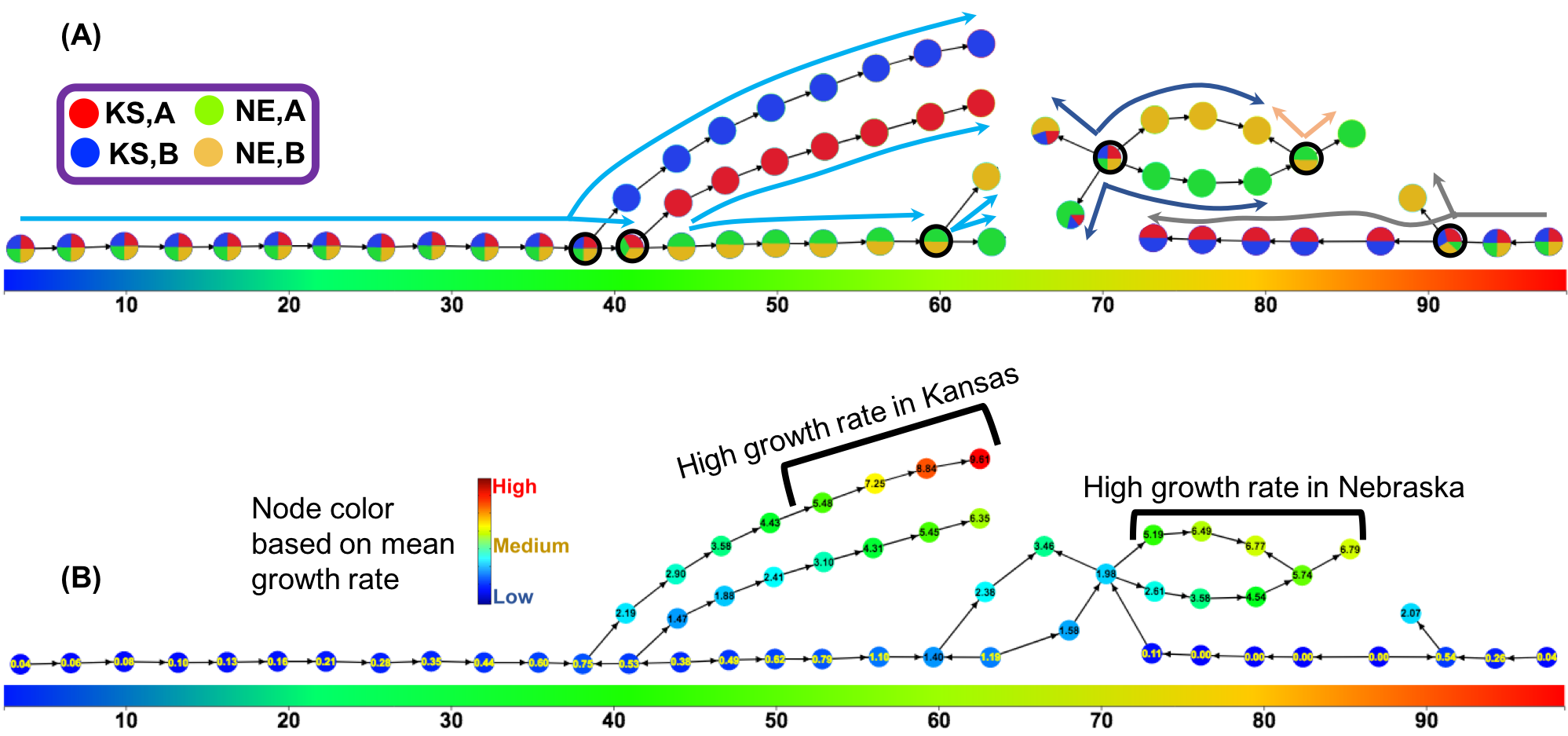}
%\medskip
\caption{
Topological object constructed using DAP as a single filter function and our method detected interesting flares from the object.
The horizontal color bar indicates the gradient of DAP, with the value increasing from left to right.
(A) Each cluster (node) of the topological object is rendered as a pie-chart showing the distribution of their four classes of individuals. 
Long arcs of different colors show interesting flares, and the corresponding branching nodes are identified with bold border. 
The blue flare (long arc spanning DAP $1$ through $60$) was ranked as the top interesting flare. 
(B) Each cluster colored by its mean growth rate (phenotype), with branches showing active growth (high phenotype) marked.
}
\label{fig:1D_flare}
\end{figure}

The resulting object along with the detected flares are shown in \figurename~\ref{fig:1D_flare}, based on which we make the following observations.
\begin{enumerate}
\item Until around DAP $\sim$40, all four subpopulations behave similarly (as shown by the leading trail of clusters). 
\item Around DAP $\sim$40, two branching events emerge:
  i) The first branching event occurs when the \{KS,B\} subpopulation separates from the rest due to a significantly accelerated growth spurt (compared to the rest).
  ii) The second event corresponds to the \{KS,A\} subpopulation separating from the rest.
  \figurename~\ref{fig:1D_flare}(B) shows the cluster nodes colored by growth rate.
\item It is not until DAP $\sim$70 that the Nebraska varieties show a similar separation in their behavior. 
\end{enumerate}

All the above branching events were successfully detected by our flare detection algorithm (shown by long arcs of different colors) in a runtime of $9$ milliseconds after the Mapper graph is 
built.
The runtime to construct mapper graph from the KS/NE data set was $167$ milliseconds.
Note that our method is \emph{unsupervised}---the information about the source genotypes and locations (pie-chart distribution in \figurename~\ref{fig:1D_flare}(A)) was applied only \emph{after} the analysis was completed, just to aid in our interpretation. 
These results demonstrate our method's ability to successfully delineate interesting
subpopulations that show divergent behavior in an unsupervised manner.

Our path detection algorithm also identified interesting paths in the 
object of \figurename~\ref{fig:1D_flare} which are already covered by either a stem of a flare or a branch of a flare 
or both.
Therefore, the observations we can make based on paths using the single filter function DAP
are similar to ones we made using flares.

\subsection{Two filter functions}\label{sec:dfFltrFnc} 
In the results for single filter function given above,
the fact that genotype B in Kansas shows a significantly altered behavior
compared to the same genotype in Nebraska indicates that there could be causal
environmental factors at play that influence the phenotype. To better
characterize such potential candidates for key environmental variables, we
conduct two-filter studies (one filter being time or DAP, and another filter
being one of the many environmental variables recorded). We explored choices
of a multitude of environmental variables. In the interest of space, we present
the results for \{DAP, humidity\} combination as it led to more interesting
observations compared to other variables.

\begin{figure}[htp!]
\centering
\includegraphics[keepaspectratio=yes, width=\textwidth]{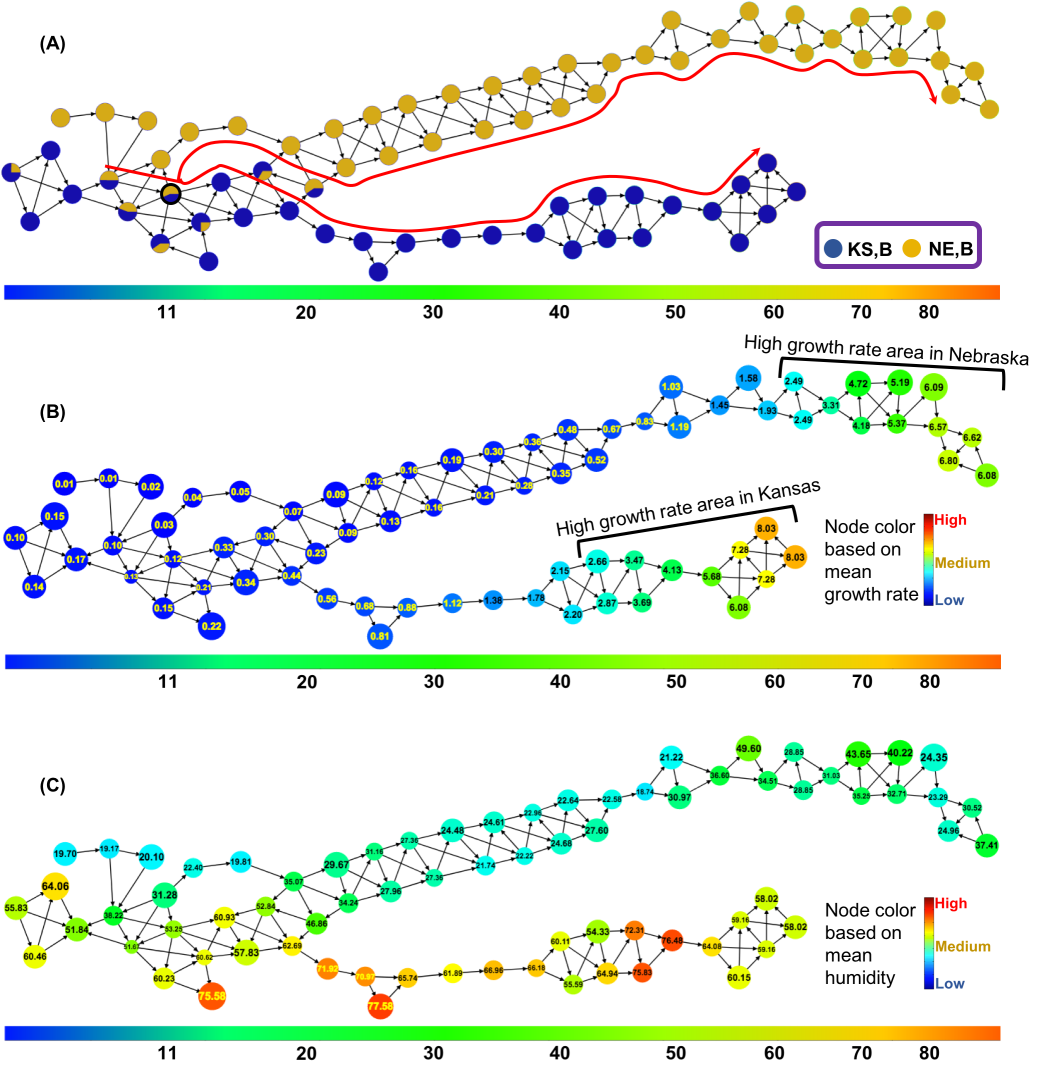}
%\medskip
\caption{
The topological object constructed using only the individuals of genotype B, using DAP and humidity as the two filter functions.
The horizontal color bar indicates the gradient of DAP, with its value increasing from left to right.
(A) Each cluster (node) is rendered as a pie-chart showing the distribution of its
individuals from the two locations (KS and NE) for genotype B. 
Parts (B) and (C) show the same topological object, however with each cluster (node)
colored by the growth rate (phenotype) and humidity (environment), respectively. 
Our method captured one large flare, which is indicated by the red branched arc in Part (A).
}
\label{fig:2D_flare}
\end{figure}

{\bf Flares: } 
\figurename~\ref{fig:2D_flare} shows the corresponding topological object on which we show
flares.
Based on this figure, we make the following observations:
%where we capture the scenario from early growth stage to mature growth stage (when growth rate is peak). We are not showing  disjoint subgraph that captures the post mature growth stage because our method did not detect meaningful flare over there.  
%Along the object the time (DAP) grows from left to right and is shown in a horizontal gradient color bar.

\begin{enumerate}
\item \figurename~\ref{fig:2D_flare}(A) shows that in the initial growth period ($1$--$10$ DAP), the performance at both locations are highly comparable, as is evidenced by the clustering of both locations. 
\item Around DAP $11$ the locations diverge into two separate branches (as shown in panel (A)). 
  This separation is correlated with variation in local humidity values (see panel (C))---more specifically, while Nebraska experienced steadily low humidity values until around DAP $50$, Kansas experienced fluctuating and often high humid conditions for most of the period until around DAP $60$. 
  This period of high humidity fluctuation also coincides with the accelerated growth rate that Kansas experiences from around DAP $40$ (panel (B)). 
  As for Nebraska, the increase in growth rates occur eventually around DAP $60$ (panel (B)) and that too coincides with higher values in humidity (panel (C)).
\end{enumerate}

{\bf Path analysis: } 
In the next step, we ran our interesting path detection algorithm, as described in
Section~\ref{sec:IP}.
All runs were performed with the following settings:
i)  each path detected is such that all its edges have the same signature; and
ii) each path should have at least $3$ edges. 

The collection of paths that were detected by our algorithm roughly divide the topological object (\figurename~\ref{fig:fig_path_B}) into three growth stages at each location (Kansas/Nebraska) based on the growth rate of plants with respect to the time (DAP).
These growth stages are a) Early growth stage, b) Mid-growth stage, and c) Mature growth stage, as shown in \figurename~\ref{fig:fig_path}.
We describe the growth stages and related observations in more detail.

%We now evaluate the qualitative significance of the interesting paths identified by \hyppox{}, shown in \figurename~\ref{fig:fig_path}:

\begin{itemize}
  \item {\bf Early growth stage:}
    The collection of co-located paths $P_7, P_{10}, P_{11}, P_{12}$ helps us understand how the genotype behaves in its early stages of development in the two locations.
    More specifically, both paths [$P_7,P_{11}$] capture nodes that contain points from both locations because their performances in similar conditions (DAP and humidity) are also quite similar;
    however, after roughly $22$ days after planting (\figurename~\ref{fig:fig_path_C}), the points from KS and NE separate (into $P_{10}$ and $P_{12}$ respectively). 
  
  \item {\bf Mid-growth stage in Kansas:}
    The sequence of paths [$P_6,P_5,P_1$], which also includes the \emph{most} interesting path by interestingness score ($P_1$), represents the active growth period for the KS population (see \figurename~\ref{fig:fig_path_B}). 
    In this period, the growth rate increased from $1.38$ cm/day to $8.03$ cm/day, from approximately $35$ days after planting to $61$ days after planting (see \figurename~\ref{fig:fig_path_C}). 
    In contrast, the plants in NE, despite being the same genotype, had very low growth rates during roughly the same period in time ($39$ days after planting to $60$ days after planting; see paths [$P_{9},P_8$] of both Figs.~\ref{fig:fig_path_B} and \ref{fig:fig_path_C}).

    \smallskip
    Incidentally, examining the humidity trends in the same period for these two locations (see \figurename~\ref{fig:fig_path_D}), we see that the humidity was very low in NE compared to KS,
    and that the increase in humidity values for the NE population (after $56$ days after planting) coincides with the increased activity in its growth rate (see Figures~\ref{fig:fig_path_C} and \ref{fig:fig_path_D}).
    This observation gives us an indicator that humidity may have an active role in NE, perhaps more so than in KS, in accelerating growth rate during the mid-stages of development.
  
  \item {\bf Mid-growth stage in Nebraska:}
    The sequences of paths [$P_9,P_8$] and [$P_3,P_4$] represent the active growth period of the NE population (more specifically, the growth burst starts from the middle of the path $P_8$), where the growth rate increases from $1.19$ cm/day to $6.57$ cm/day (\figurename~\ref{fig:fig_path_B}). 
    This high activity period starts from approximately $56$ days after planting and ends roughly at $80$ days after planting.
    As indicated above, this active growth rate coincides with the period having higher humidity for NE.

  \item {\bf Mature growth stage: }
    The path $P_2$ helps us understand how the genotype behaves in its later stages of development in the two locations.
    More specifically, path $P_2$ starts with points from both locations because their performances in similar conditions (DAP and humidity) are also quite similar.
    However, after roughly $92$ days after planting (\figurename~\ref{fig:fig_path_C}),  plants in both locations do not grow much. 
	
  \item
    The path $P_7$ and $P_2$ illustrate that plants of the genotype $B$ do \emph{not} grow as much before $22$ days after planting and after $92$ days after planting, respectively. 

\end{itemize}

\begin{figure}[hbp!]
  \centering
  \begin{subfigure}[htp!]{\textwidth}
    \renewcommand*{\thesubfigure}{\Alph{subfigure}}
    \includegraphics[keepaspectratio=yes, width=\textwidth]{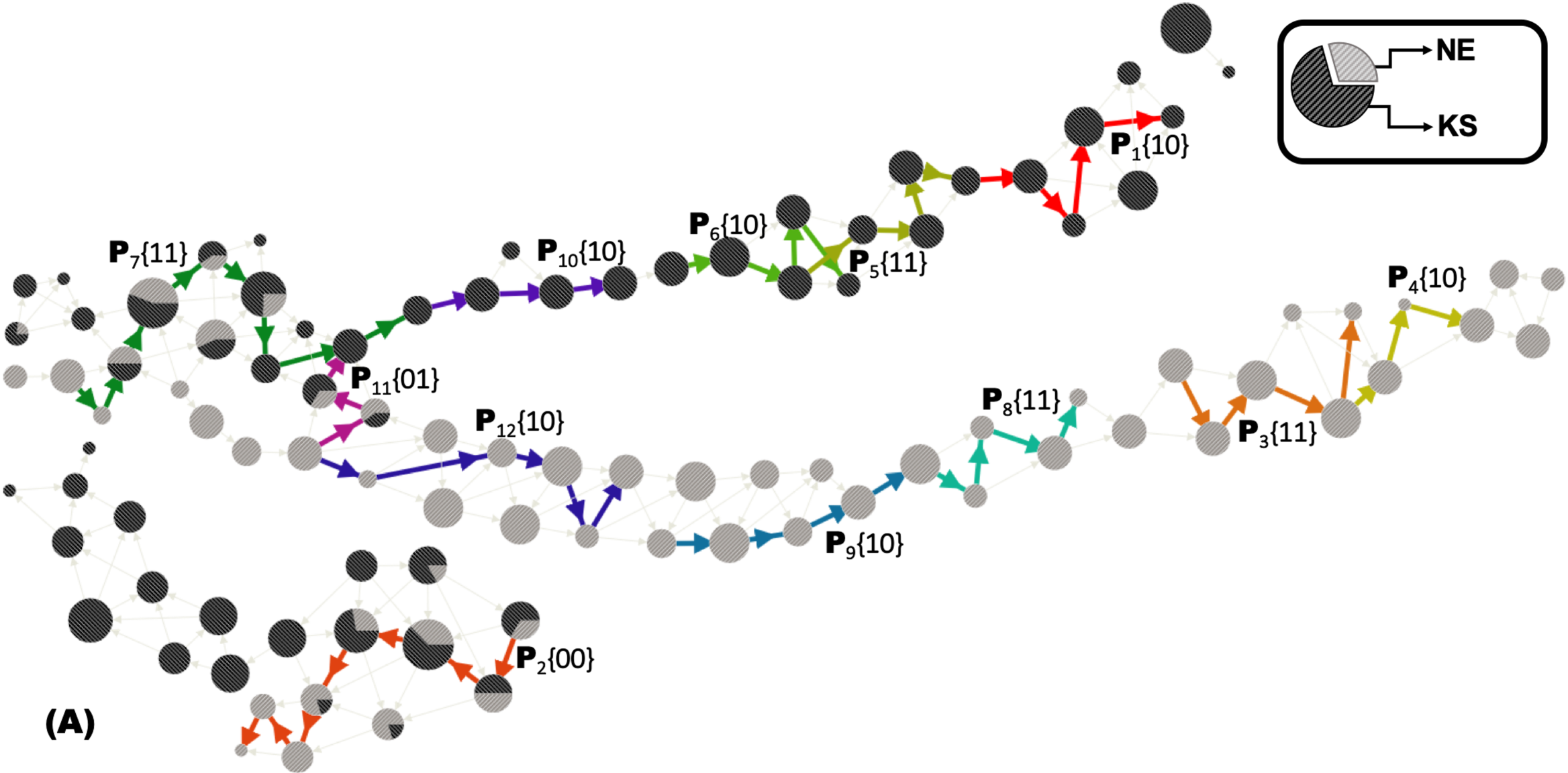}
    \subcaption{Relative concentrations of locations (KS and NE) in each cluster.  \medskip}
    \label{fig:fig_path_A}
  \end{subfigure}
  \begin{subfigure}[htp!]{\textwidth}
    \renewcommand*{\thesubfigure}{\Alph{subfigure}}
    \includegraphics[keepaspectratio=yes, width=\textwidth]{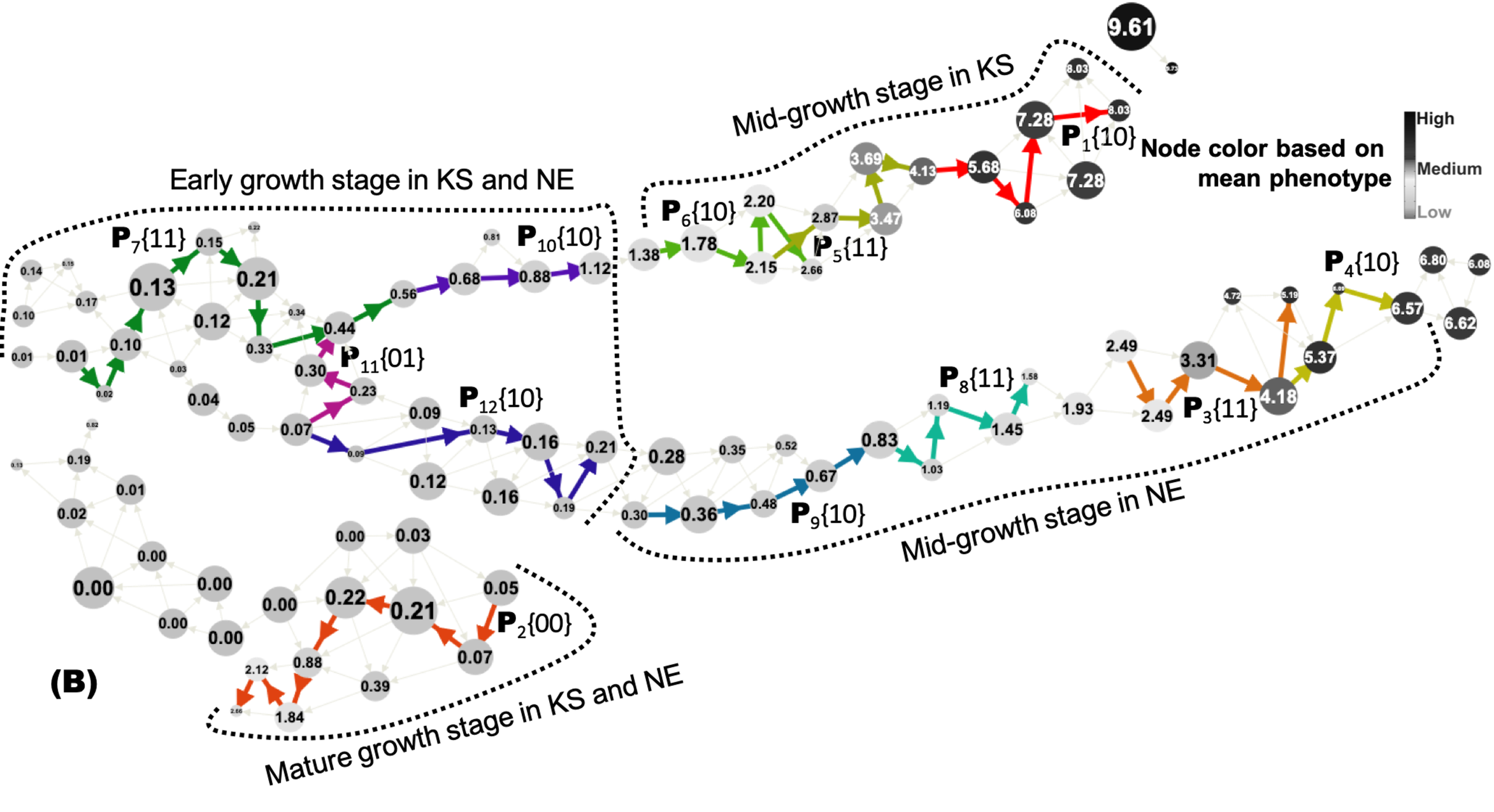}
    \subcaption{Mean phenotypic value for each cluster, with shade going from light gray (low) to black (high).}
    \label{fig:fig_path_B}
  \end{subfigure}
  \caption{Topological object constructed using DAP and humidity as the two filter functions, using only the [KS,B] and [NE,B] points.
    Figure is continued on the next page---see Caption for details.
    }
\end{figure}

\begin{figure}[htp!]\ContinuedFloat
  \centering
  \begin{subfigure}[htp!]{\textwidth}
    \renewcommand*{\thesubfigure}{\Alph{subfigure}}
    \includegraphics[keepaspectratio=yes, width=\textwidth]{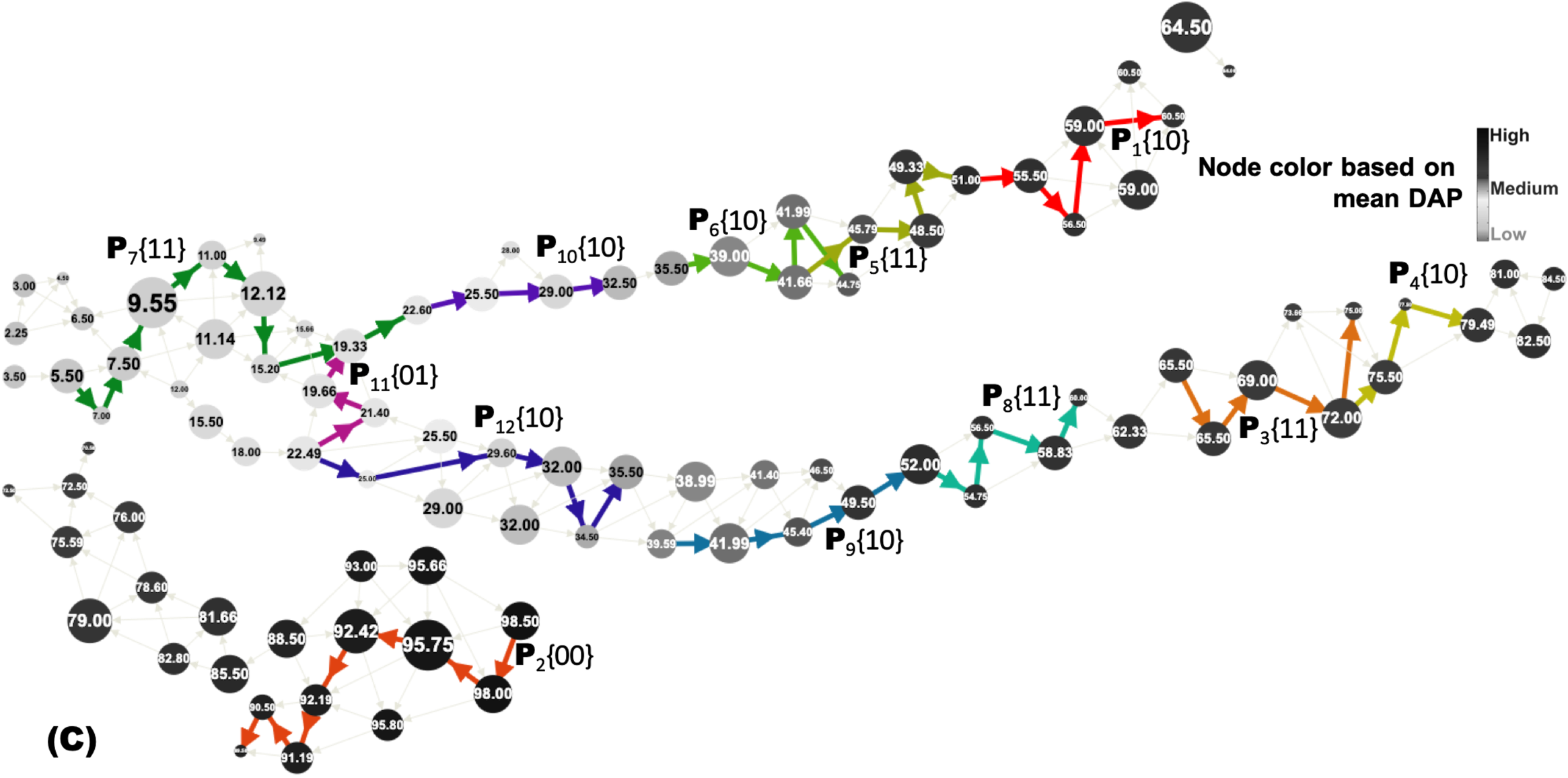}
    \subcaption{Mean DAP value for each cluster, with shade going from light Gray (low) to black (high).}
    \label{fig:fig_path_C}
  \end{subfigure}
  \begin{subfigure}[htp!]{\textwidth}
    \renewcommand*{\thesubfigure}{\Alph{subfigure}}
    \includegraphics[keepaspectratio=yes, width=\textwidth]{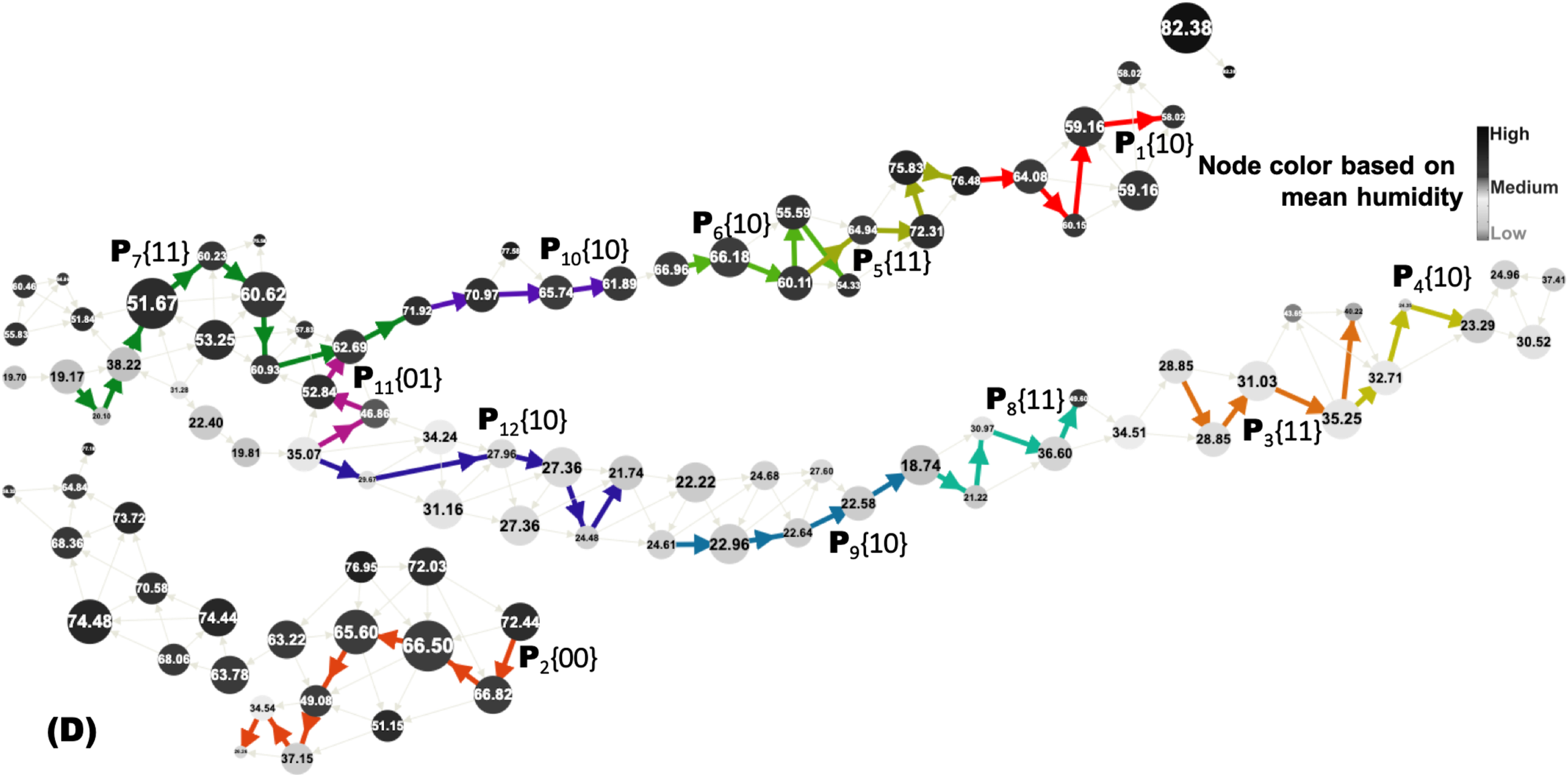}
    \subcaption{Mean humidity value for each cluster, with shade going from light Gray (low) to black (high).}
    \label{fig:fig_path_D}
  \end{subfigure}
  \caption{
    Topological object constructed using DAP and humidity as the two filter functions, using only the [KS,B] and [NE,B] points.
    The same object is shown in the four panes, albeit with different gray scale coloring schemes. 
    Figure (\ref{fig:fig_path_A}) shows each node (i.e., partial cluster) as a pie-chart of the relative concentrations of the two possible locations combinations;
    (\ref{fig:fig_path_B}) shows nodes colored by their mean phenotypic value;
    (\ref{fig:fig_path_C}) shows nodes colored by their mean DAP value; and
    (\ref{fig:fig_path_D}) shows nodes colored by their mean humidity value.
    The mean values are also indicated within the respective circles. 
    The size of the circle for each node is proportional to the size of the corresponding partial cluster.
    Also shown highlighted as thick colored edges are the set of interesting paths identified by our method. 
    Edge directions are from low to high mean phenotypic values.
    The interesting paths are labeled as $P_i\{s_1s_2\}$, where $i$ is the path number, and $s_1s_2$ denotes the signature for that path ($s_1$ corresponds to DAP and $s_2$ corresponds to humidity). 
    Recall that in the signature, $0$ means decreasing and $1$ means increasing.
  }
  \label{fig:fig_path}
\end{figure}

\clearpage
To better understand the results of \hyppox{} and contrast the 
capability of our method with more traditional approaches, 
we plotted all the genotype $B$ points as a scatter plot, based
on their  DAP and humidity (see \figurename~\ref{fig:fig_2D_path}).
The coloring of the points are by their location. 
As can be seen, the plot shows a clear separation between NE and KS
humidity values, with NE exposed to lower humidity values than KS, in
general. Note that this is a coarse-level information which could have been easily
obtained through a correlation test as well. However the limitation of 
such correlation tests is that they point to global trends without providing
insight into the variabilities that may exist across different subpopulations at
different scales. 
On the other hand, identifying such subpopulation-based variability 
(as output by the paths and flares from \hyppox)
could prove useful in delineating key environmental or temporal triggers that impact crop performance, 
and on how that behavior varies within a diverse population.
That is where our topology-based approach can be useful---to make such
inferences from the data and formulate testable hypotheses.

To better illustrate this advantage, we overlaid the interesting path sequences
identified by our paths (discussed above) on to the scatter plot. These
path sequences are shown as arcs in \figurename~\ref{fig:fig_2D_path}.  
As can be seen, our interesting paths show four major ``features'' within
this scatter plot:
\begin{compactenum}[i)]
\item
  the initial sequence where both NE and KS varieties behave similarly in their initial developmental stages, before branching out (around $22$ DAP);
\item
  the period of active growth for [KS,B] between roughly $35$ and $61$ days after planting; 
\item 
  the period of active growth for [NE,B] appearing much later, between roughly $56$ and $80$ days after planting; and
\item 
  the observation of the path that separates at around $22$ days after planting merging back after $92$ days after planting.
\end{compactenum}

More interestingly, at the end of our interesting paths ([$P_9,P_8$]) for [NE,B] is also for the first time the humidity value experienced a spike for that location---increasing from values under $35$ to around $50$---effectively implying (or at least indicating) a probable cause for increased growth activity.
After that trigger, minor fluctuations in humidity seemed to have little effect in the growth rate, which continued to increase through $80$ days after planting.
This study sets up a testable link between a genotype (B) and environmental variable (in this case, humidity) toward a performance trait (growth rate). 

%Furthermore, the study raises a plausible working
%hypothesis that can be tested:
%``\emph{If genotype [NE,B] is also exposed to a higher level humidity earlier on,
%during its developmental stage, then it is also likely to show an active
%growth rate earlier}''

%\vspace*{-0.1in}
%{\bf Summary of findings:}
These results and observations suggest two plausible hypotheses:
(a) humidity is likely to influence the growth rate; and
(b) this degree of influence is more pronounced on genotype B than for genotype A. 
The precise time and humidity intervals where such effects manifest are shown by the flare. 

This illustrative example serves to demonstrate that our topology-based method also has the potential to enrich further the information that can be obtained through conventional methods such as scatter plots. 
%Note that our tool is meant for exploring high-dimensional data in a software-guided manner and more/other environmental variables can be included in our tests. 

%\input{flare_result}
%\input{path_result}
\subsection{Application on Irrigation-controlled data set}\label{sec:irr-result}
\subsubsection{Irrigation data set}\label{sec:data2}
%\add{
This maize data set consists of phenotypic and environmental measurements for $80$ genotypes, grown in two field locations in Nebraska (NE), USA. 
Over the growing season, one field location solely depended on rainfall whereas irrigation facility was provided to the other field location. 
Apart from this irrigation facility, all other environmental parameters are identical 
in both field locations.
The data consists of \emph{daily} measurements of the genotypes' growth rates alongside multiple environmental variables, over the course of the first $80$ days of the growing season.
For the purpose of our analysis we treat each unique [genotype, time] combination as a ``point''. 
For each point, we computed the growth rate difference from the irrigated location to 
the non-irrigated location.
Consequently, the above data set consists of $N=6400$ points.
Here, ``time'' was measured in Days After Planting (DAP). 
An ``individual'' in this data set refers to a specific genotypic plant.
%}

%\add{
%Each point has one phenotypic value (growth rate difference) and $8$ environmental variables (including but not limited to: humidity, temperature, rainfall, and solar radiation).  We studied the choice of a multitude of environmental variables; In the interest of space, we present the results for DAP here, as it led to more interesting observations compared to the other variables.}

\begin{figure}[hbp!]
  \centering
  \includegraphics[keepaspectratio=yes, width=\textwidth]{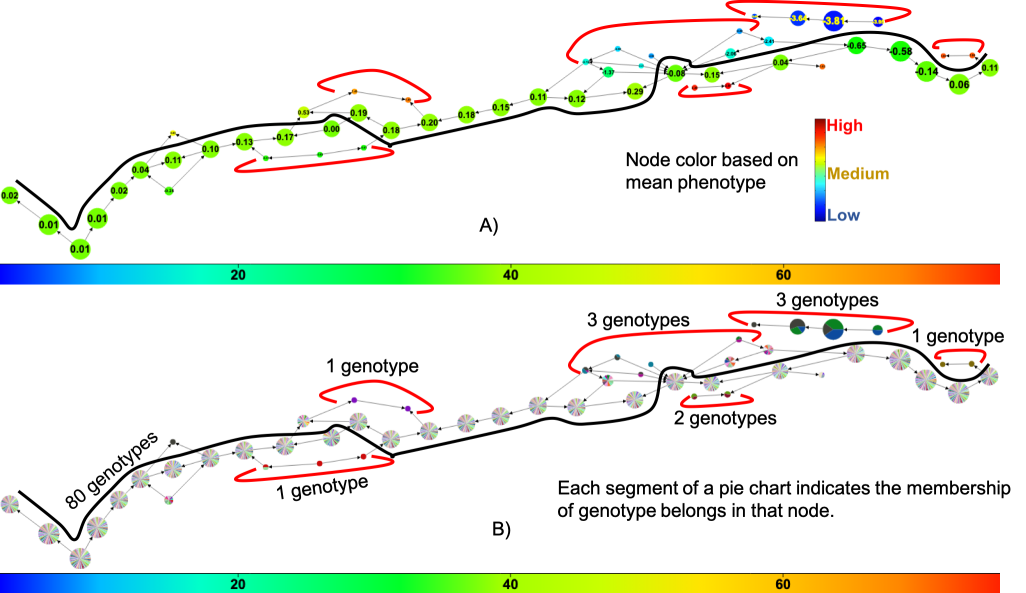}
  \medskip
  \caption{
    Topological object constructed using DAP as a single filter function.
    The horizontal color bar indicates the gradient of DAP, with the value increasing from left to right.
    All the nodes along black colored arc shows almost similar growth rate 
    difference (see part (A)) which means the growth rate difference between irrigated and non-irrigated locations for all the genotypes (see part (B)) that belong
    in this path are very close to each other. We also observed some nodes (marked by red-colored arcs) 
    with divergent phenotypic values, which indicate that the genotypes (see part (B))
    of these nodes have phenotypic variation between two different controlled 
    environments (irrigated and non-irrigated). A detailed overview of these genotypes 
    are given in Appendix~\ref{sec:vis-plastic}.
  }
  \label{fig:LD_Plastic}
\end{figure}

{\bfseries Topological object construction:}
%\add{
First, we constructed our topological object using DAP as a single filter function
(parameter setting for this analysis is given in Table~\ref{tbl:LD_DAP})
and phenotypic difference between points for clustering.
This study is aimed at understanding how the population of individuals (genotypes) show varying trends in phenotypic performance (i.e., growth rate here) as a function of time with respect to two distinct controlled environments (irrigated and non-irrigated).
The resulting object is shown in \figurename~\ref{fig:LD_Plastic}.
%}

\medskip
\begin{table}[htp!]
\renewcommand{\arraystretch}{1.3}
\caption{Parameter settings for single filter (DAP) analysis on irrigation-controlled dataset.}
\label{tbl:LD_DAP}
\centering
\begin{tabular}{|c|c|}
\hline
\bfseries Steps & \bfseries Parameters\\\hline
Filtering & $27$ windows along filter DAP \\\hline
Clustering & $r=0.2$, $\rho = 2$ \\\hline
Persistent overlap & $20\%$ \\\hline
\end{tabular}
\end{table}

\bigskip

\clearpage
{\bfseries Object exploration:}
%\add{
From the resulting topological object shown in  \figurename~\ref{fig:LD_Plastic}, we observed phenotypic 
variation of some of the genotypes when they were exposed in 
two different controlled environments (irrigated and non-irrigated).
According to \figurename~\ref{fig:LD_Plastic}(A), all the nodes along the bold black 
line show similar performance, which indicates that the points belonging to these nodes 
have small growth rate variation between irrigated and non-irrigated 
environments. On the other hand, the nodes those are marked by  red-colored 
arcs contain points which show large growth rate variation between irrigated 
and non-irrigated environments. The genotypes listed in Table~\ref{tbl:LD_plastic} 
are retrieved from the points belonging to the nodes marked by red arcs (\figurename~\ref{fig:LD_Plastic}(B)).
%}

\bigskip
\begin{table}[htp!]
\renewcommand{\arraystretch}{1.3}
\caption{List of root worm affected genotypic plants and their corresponding DAP range.}
\label{tbl:LD_plastic}
\centering
\begin{tabular}{|c|c|c|}
\hline
\bfseries Genotype & \bfseries Starting DAP & \bfseries Ending DAP\\\hline
PHW52 x LH123HT & $27$ & $36$ \\\hline
PHB47 x PHR55 & $33$ & $39$ \\\hline
LH198 x PHW30 & $54$ & $63$ \\\hline
PHW52 x Q381 & $51$ & $57$ \\\hline
PHP02 x PHB47 &  $59$ & $63$ \\\hline
PHB47 x LH185 & $57$ & $63$ \\\hline
PHB47 x PHG83 & $51$ & $67$ \\\hline
LH198 x LH51 & $62$ & $69$ \\\hline
PHB47 x LH38 &  $66$ & $69$ \\\hline
ICI 441 x PHZ51 & $74$ & $78$ \\\hline
\end{tabular}
\end{table}

\medskip
%\add{
The set of genotypes covered in the  nodes marked by red arcs are interesting because each of them shows phenotypic variation when exposed to two environments.
%All the attributes of the two exposed environments are very identical rather than the irrigation condition because both fields are very close to each other. Therefore, each of the genotypes should produce same growth-rate trand at each location. Hence, the growth-rate difference between two locations should be similar for all the genotypes. 
Originally, our working hypothesis was that this phenotypic divergence is a result of irrigation vs.~non-irrigation, or some other factors that affected growth in these plants. 
Upon careful examination by domain scientists who generated the data (Hey and Schnable, coauthors of this manuscript), we indeed confirmed that the plants selected in these genotypes were affected by a root-worm disease. 
Root-worm is an insect that cuts roots of a plant, which leads to the death of the plants. 
Our method is able to detect such affected genotypic plants in an unsupervised way. 
For a closer look, we zoomed into part of the topological object (\figurename~\ref{fig:LD_Plastic}(B))
and marked the starting and ending DAP for each of the genotypes listed in Table~\ref{tbl:LD_plastic} in Appendix~\ref{sec:vis-plastic}.
%}

%{\bf Different kinds of divergent behavior:}
%\add{
The above application on the irrigation controlled data set also shows the ability of our framework to extract another type of topological feature---one of \emph{spines} where majority of the points follow one behavioral pattern while a small subset of points deviate (as divergent paths).
For instance, in  \figurename~\ref{fig:LD_Plastic} spines are indicated by the thick black arc, while the divergent paths are identified by red arcs. 
An extension of our framework could be to score and identify interesting spines similar to how we identified flares and paths. 
Another related extension is one of finding highly traversed paths in the topological object and compare them to less traversed paths. 
%}

%In addition flares and paths, our framework can also be used to mine for divergent subpopulations which manifest themselves as \emph{spines} within a certain interval of time. This is generally applicable to data sets where majority of the points follow one trend and there are few divergent paths. To demonstrate one such use case, we used a real world application. 
%This can be further formularized into a scoring scheme similar to flare scoring model.

%\add{In a similar manner described for KS/NE data set, we can also find flares and paths, however, given the nature of the particular data set the directionality in the generated graph needs to be ignored in order to find heavily traverse paths as suppose to lesser traverse paths.  }

\section{Conclusion}\label{sec:conclusion}
We have presented a scalable exploratory framework for navigating
high-dimensional data sets and applied it to plant phenomics data to 
analyze the effect of environmental factors on phenotypic traits.
At its core, our approach is fundamentally different from state-of-the-art 
techniques in many ways as outlined below.
First, it inherits the advantages of topology including its use of 
coordinate-free representations, robustness to noise, and natural rendition of 
compact representations.
Second, by allowing the user to define multiple filter functions, it enables
them to study the combined effect of multiple factors on 
target performance traits.
Third, through its clustering and visualization capabilities, it provides a
way for domain experts to readily observe emergent behavior among different
groups or subpopulations without requiring the knowledge of any priors.
This feature enables scientists to identify subpopulations, compare them, 
and perform more targeted studies to formulate and test hypotheses.

Our approach is scalable in that it can scale to large data sets containing
possibly tens of thousands of points, reducing such large data to tens or
hundreds of partial clusters, thereby making visualization and exploration
possible. 
%Although we have presented  results on a smaller data set, we have tested our approach on larger data sets (e.g., with thousands of points \cite{syngenta}; we did not present these results due to some missing information about the data (e.g., genotypes)).

While the scope of this work can be further expanded through application to a
broader range of phenomics data collections, the results presented in this
paper show a promising application of topology and its role in hypothesis
extraction from high-dimensional data sets.
Considering the nascency of the phenomics field, tools for users to
explore data and help extract plausible hypothesis in a
data-guided manner from large-scale complex data will be important going forward.

\section*{Acknowledgment}

The research was supported by U.S. National Science Foundation grant DBI 1661348. 

%\bibliographystyle{plainurl}
%\bibliography{ABI_2015,BigData_nsf14,Zhiwu_Zhang_refs_FIX,homology,Medicine,AK_paper,Refs_Prot}

\clearpage
%\newpage
%\appendices
\appendix
\noindent {\huge {\bfseries Appendices}}

\section{Root worm affected genotypes}\label{sec:vis-plastic}
\begin{figure}[htp!]
\centering
\includegraphics[keepaspectratio=yes, width=\columnwidth]{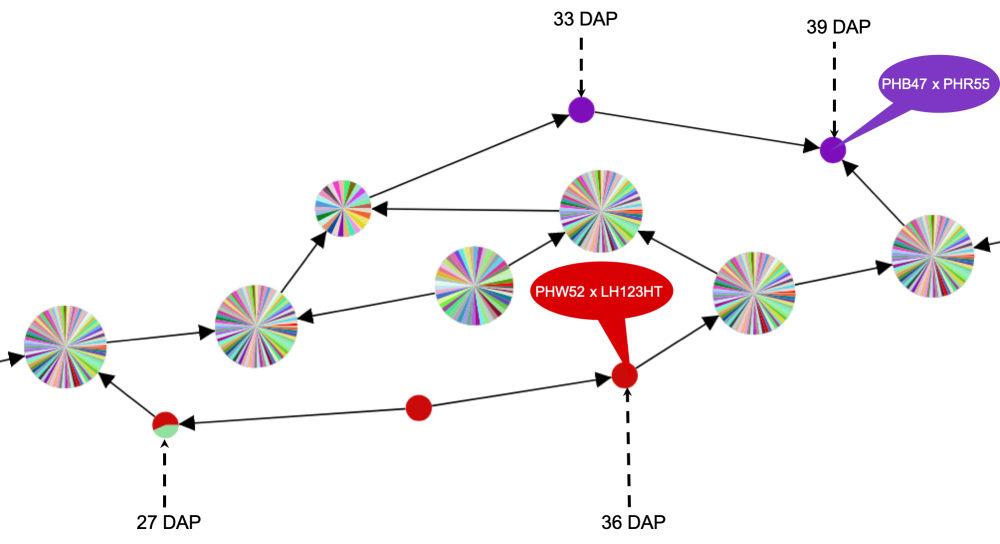}
\caption{
Genotype ``PHW52 x LH123HT'' shows phenotypic variation from $27$ DAP to $36$ DAP 
whereas, genotype ``PHB47 x PHR55'' shows phenotypic variation from $33$ DAP to $39$ 
DAP.
} 
\label{fig:LD_gen_12}
\end{figure}

From the topological object of \figurename~\ref{fig:LD_Plastic}(B),  we visually identified all the genotypes those are listed in 
Table~\ref{tbl:LD_plastic}. All the genotypes of this table are affected by root 
worm disease, that led the affected plants to die. The DAP range in the table indicates a tentative time frame when plants were affected by this disease .
To clearly observe the time duration when a genotype 
shows performance variation, we zoomed the corresponding portion of the graph. 
The zoomed portion of the main graph with a short description for each of the 
genotypes of Table~\ref{tbl:LD_plastic} are as follows:

{\bf PHW52 x LH123HT \& PHB47 x PHR55}: From \figurename~\ref{fig:LD_gen_12} we observed that 
genotype ``PHW52 x LH123HT'' shows phenotypic variation from $27$ DAP to $36$ DAP. Similarly, genotype 
``PHB47 x PHR55'' shows phenotypic variation from $33$ DAP to $39$ DAP.

\begin{figure}[htp!]
\centering
\includegraphics[keepaspectratio=yes, width=\columnwidth]{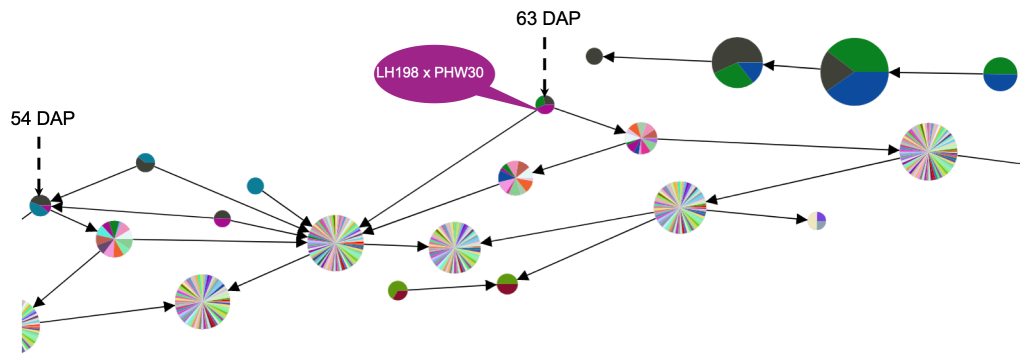}
\caption{Genotype ``LH198 x PHW30'' shows phenotypic variation between $54$ DAP to $63$ DAP.}
\label{fig:LD_gen_3}
\end{figure}

{\bf LH198 x PHW30}: From \figurename~\ref{fig:LD_gen_3} we observed that 
genotype ``LH198 x PHW30'' shows phenotypic variation from $54$ DAP to $63$ DAP. 

\begin{figure}[htp!]
\centering
\includegraphics[keepaspectratio=yes, width=\columnwidth]{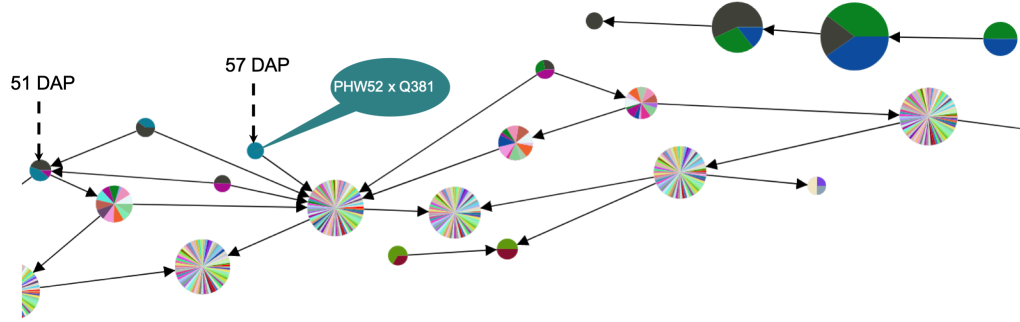}
\caption{Genotype ``PHW52 x Q381'' shows phenotypic variation between $51$ DAP to $57$ DAP.}
\label{fig:LD_gen_4}
\end{figure}

{\bf PHW52 x Q381}: From \figurename~\ref{fig:LD_gen_4} we observed that 
genotype ``PHW52 x Q381'' shows phenotypic variation from $51$ DAP to $57$ DAP. 

\begin{figure}[htp!]
\centering
\includegraphics[keepaspectratio=yes, width=\columnwidth]{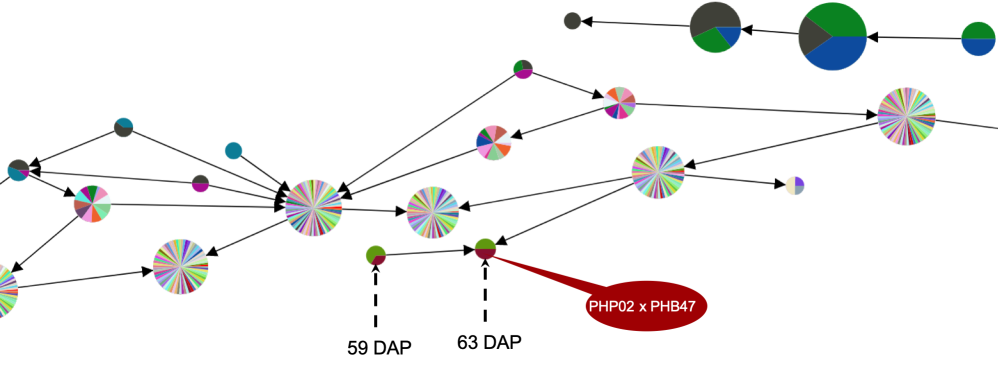}
\caption{Genotype ``PHP02 x PHB47'' shows phenotypic variation between $59$ DAP to $63$ DAP.}
\label{fig:LD_gen_5}
\end{figure}

{\bf PHP02 x PHB47}: From \figurename~\ref{fig:LD_gen_5} we observed that 
genotype ``PHP02 x PHB47'' shows phenotypic variation from $59$ DAP to $63$ DAP. 

\begin{figure}[htp!]
\centering
\includegraphics[keepaspectratio=yes, width=\columnwidth]{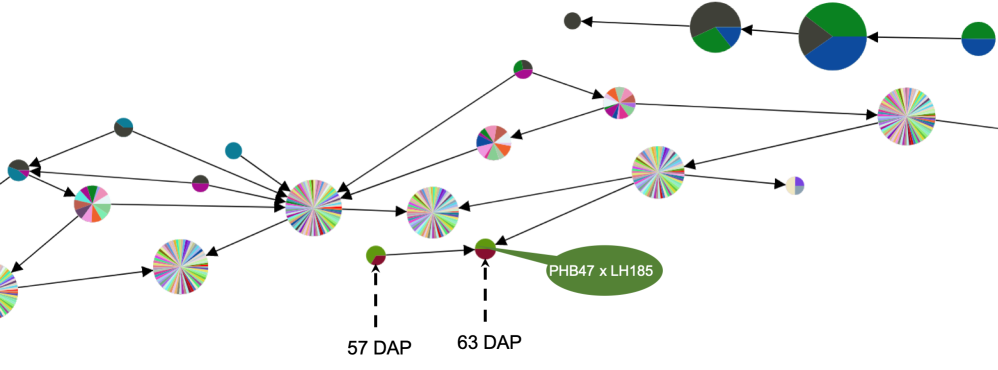}
\caption{Genotype ``PHB47 x LH185'' shows phenotypic variation between $57$ DAP to $63$ DAP.}
\label{fig:LD_gen_6}
\end{figure}

{\bf PHB47 x LH185}: From \figurename~\ref{fig:LD_gen_6} we observed that 
genotype ``PHB47 x LH185'' shows phenotypic variation from $57$ DAP to $63$ DAP. 

\begin{figure}[htp!]
\centering
\includegraphics[keepaspectratio=yes, width=\columnwidth]{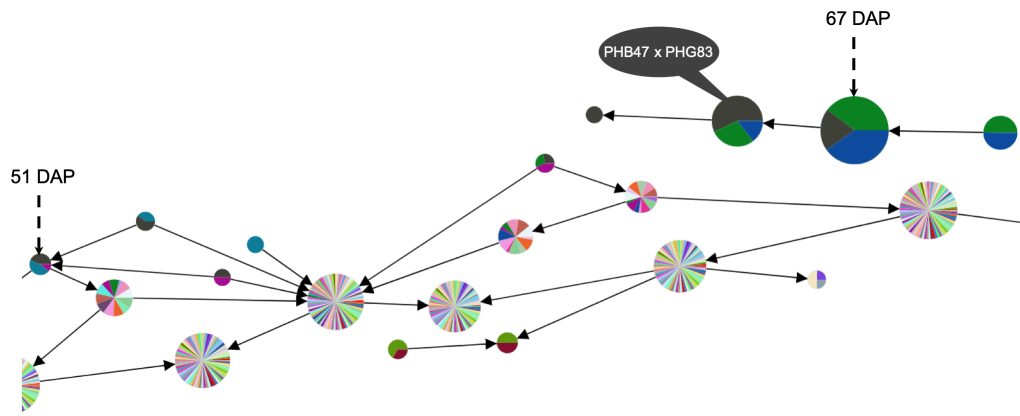}
\caption{Genotype ``PHB47 x PHG83'' shows phenotypic variation between $51$ DAP to $67$ DAP.}
\label{fig:LD_gen_7}
\end{figure}

{\bf PHB47 x PHG83}: From \figurename~\ref{fig:LD_gen_7} we observed that 
genotype ``PHB47 x PHG83'' shows phenotypic variation from $51$ DAP to $67$ DAP. 

\begin{figure}[htp!]
\centering
\includegraphics[keepaspectratio=yes, width=\columnwidth]{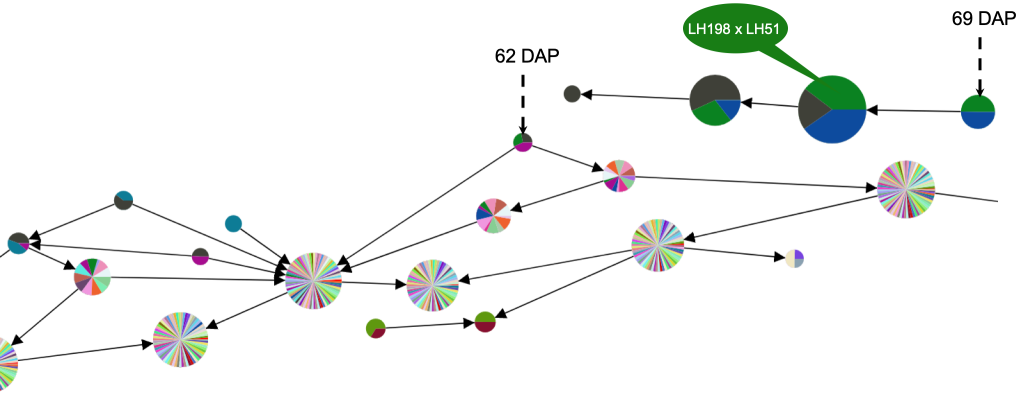}
\caption{Genotype ``LH198 x LH51'' shows phenotypic variation between $62$ DAP to $69$ DAP.}
\label{fig:LD_gen_8}
\end{figure}

{\bf LH198 x LH51}: From \figurename~\ref{fig:LD_gen_8} we observed that 
genotype ``LH198 x LH51'' shows phenotypic variation from $62$ DAP to $69$ DAP. 

\begin{figure}[htp!]
\centering
\includegraphics[keepaspectratio=yes, width=\columnwidth]{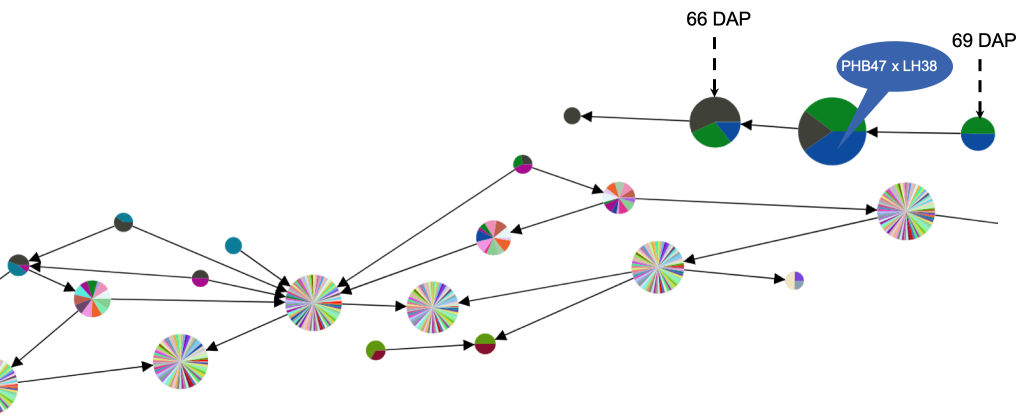}
\caption{Genotype ``PHB47 x LH38'' shows phenotypic variation between $66$ DAP to $69$ DAP.}
\label{fig:LD_gen_9}
\end{figure}

{\bf LPHB47 x LH38}: From \figurename~\ref{fig:LD_gen_9} we observed that 
genotype ``PHB47 x LH38'' shows phenotypic variation from $66$ DAP to $69$ DAP. 

\begin{figure}[htp!]
\centering
\includegraphics[keepaspectratio=yes, width=\columnwidth]{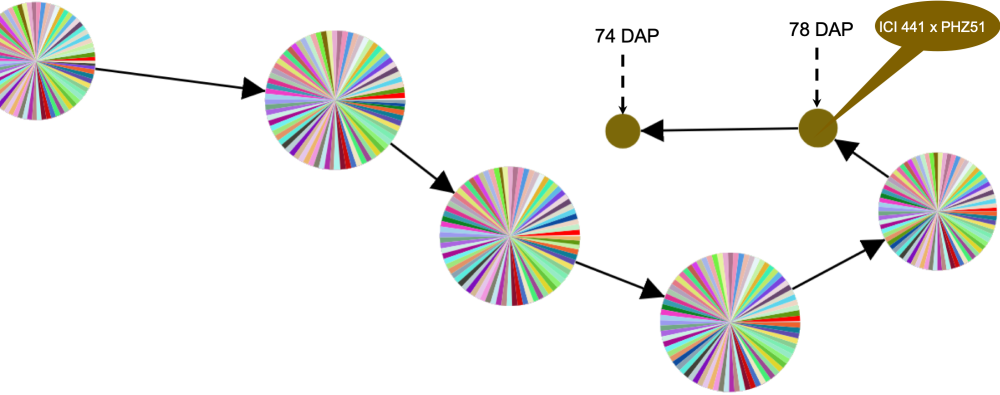}
\caption{Genotype ``ICI 441 x PHZ51'' shows phenotypic variation between $74$ DAP to $78$ DAP.}
\label{fig:LD_gen_10}
\end{figure}

{\bf ICI 441 x PHZ51}: From \figurename~\ref{fig:LD_gen_10} we observed that 
genotype ``ICI 441 x PHZ51'' shows phenotypic variation from $74$ DAP to $78$ DAP. 

\begin{figure}[htp!]
\centering
\includegraphics[keepaspectratio=yes, width=\columnwidth]{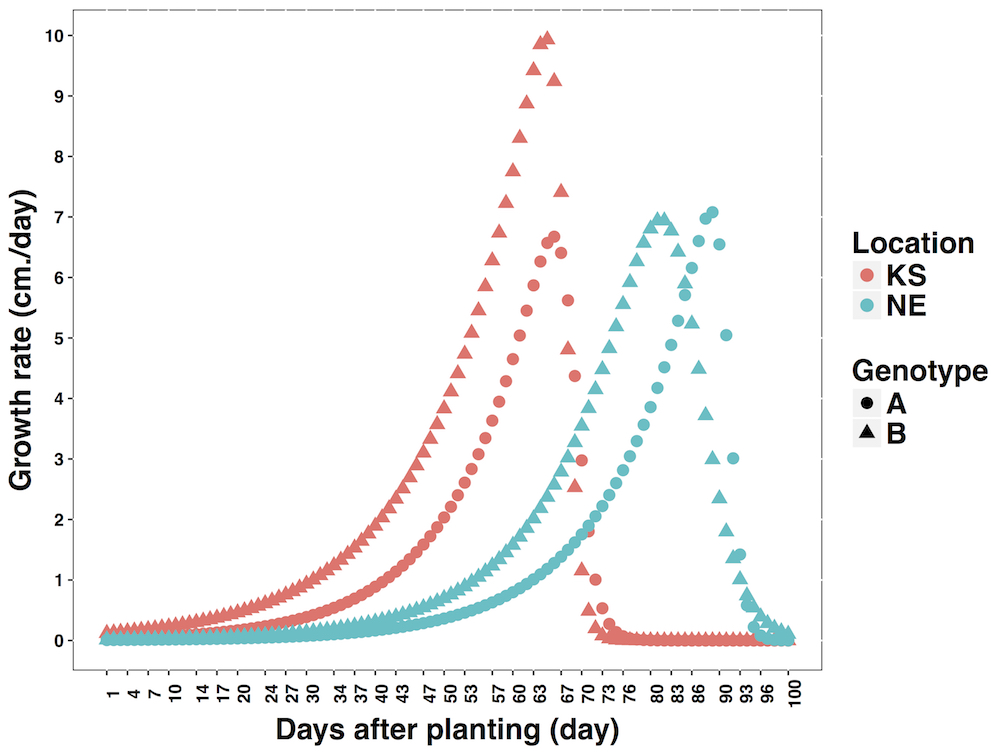}
\caption{The scatter plot of growth rate with respect to days after planting (DAP). 
Plants started to grow earlier in location $KS$ 
compared to location $NE$. Genotype $B$ of location $KS$ acts differently 
compared to both genotypes in both locations.}
\label{fig:fig_DAP_GR}
\end{figure}

\section{Edge direction in a flare}\label{sec:flare-edge-dir}
\begin{figure}[htp!]
    \centering
    \includegraphics[keepaspectratio=yes, width=3.0in]{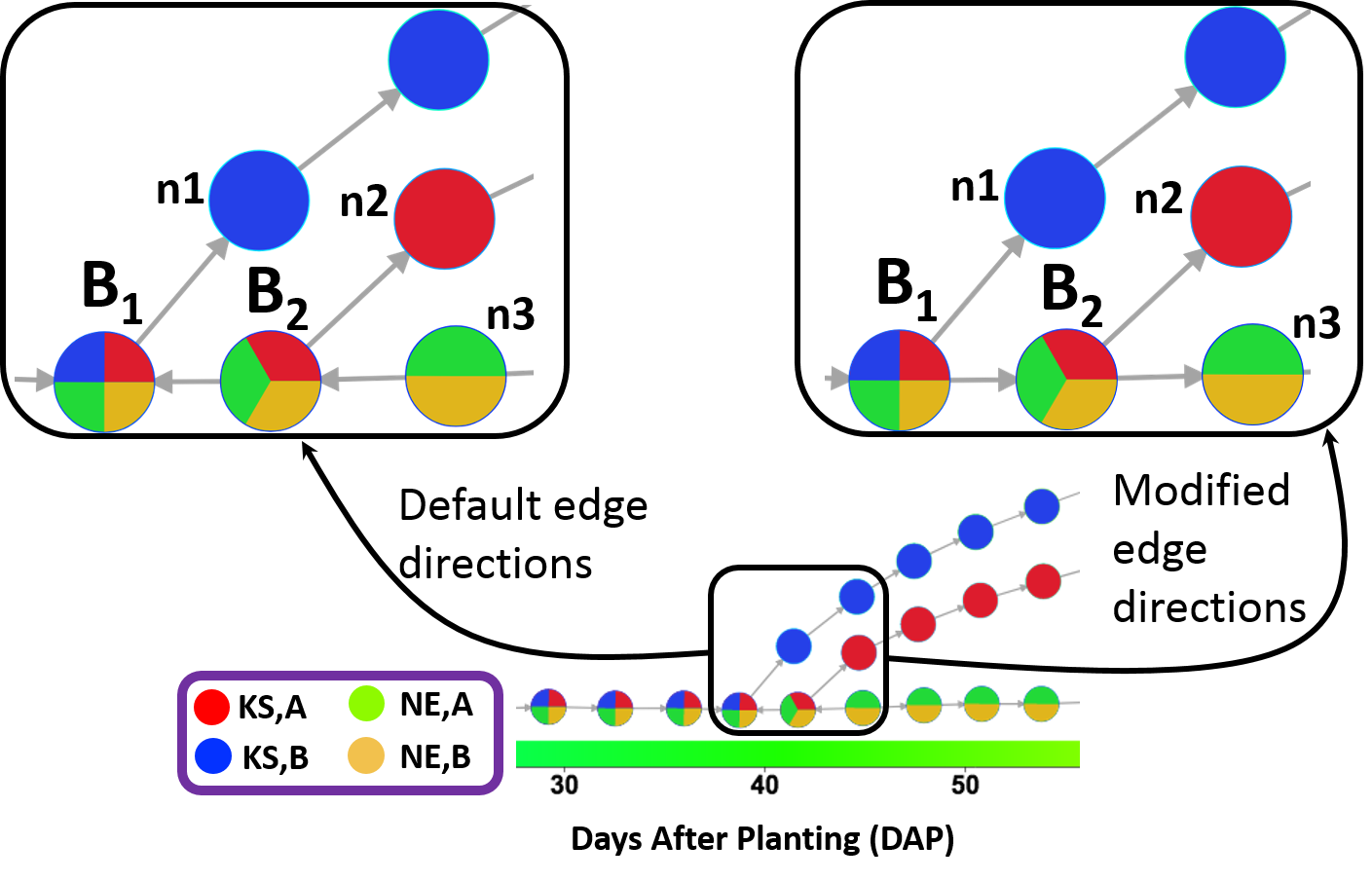}
    \caption{Modification of edge directions in the Mapper.
      Using the default approach (zoomed in on left), edges are oriented from $B_2$ to $B_1$ and from $n_3$ to $B_2$, by considering the mean phenotype values of all individuals in these nodes.
      Using the modified approach (zoomed in on right), we orient the edge from $B_1$ to $B_2$ by considering the mean phenotype value of individuals with only the genotypes shared by these nodes (i.e., (KS,A), (NE,A), and (NE,B)).
    A similar modification directs the edge from $B_2$ to $n_3$.}
    \label{fig:EdgeDir}
  \end{figure}

To capture branches in phenomics data sets accurately, we modify the way in which we direct the edges in the topological object as follows.
    Given an undirected edge $e=\{u,v\}$ in the Mapper, we direct edge $e$ from the node with the lower mean phenotypic value to the one with high value, where the respective means are now taken over the subsets of individuals in $u$ and $v$ that belong to genotypes present in both nodes.
    This procedure is illustrated in \figurename~\ref{fig:EdgeDir}.
    %In an alternative setting, we use \emph{environment} of the individuals to determine these subsets to take means over.
   % If genotype or location or environment information is not available, these means are computed over only the individuals shared by nodes $u$ and $v$.

\section{Double filter function (DAP and humidity)}
\label{sec:dap_hum}

The landscape view of \figurename~\ref{fig:3D_HUM} shows the change of phenotypic 
value with respect to both time (DAP) and environment (humidity). While some general
trends might be evident from this visualization, the question of which subset of points should be compared
to which others in order to discern the independent or combined effects of time and 
environment
cannot be answered easily, due to the explosive number of such combinations. In 
order to delineate the phenotypic variation of a subpopulation under certain 
environmental condition over a period, we applied topological data analysis 
methodology.
The step by step construction process of our 
topological object using DAP and humidity as the double filter function are as follows:

\begin{figure}[htp!]
  \centering 
   \includegraphics[keepaspectratio=yes,width=\columnwidth]{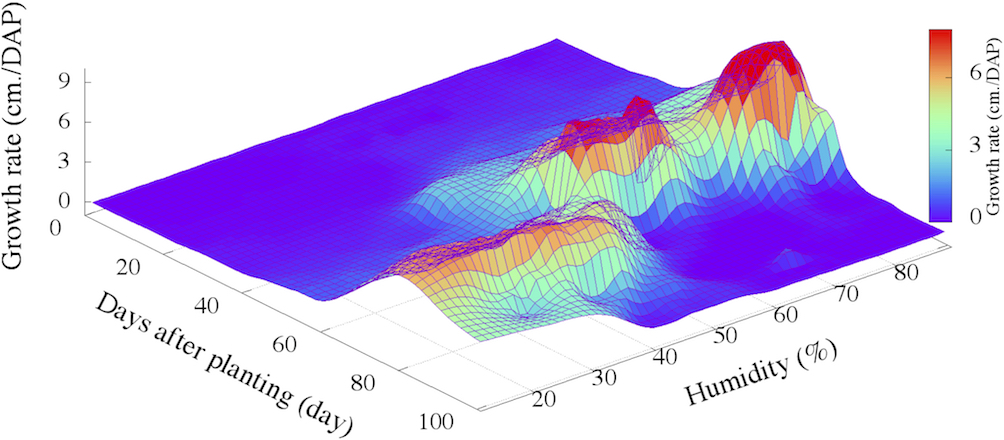}
   \caption{
3D representation of phenotype (growth rate) data with respect
to time (DAP) and one of the environmental attributes (Humidity) for a real-world
maize data set grown in Kansas (KS) and Nebraska (NE). Each point in
the landscape is a [genotype, location, date/time] combination. 
} 
\label{fig:3D_HUM} 
\end{figure}

\subsection{Filtering} 
\label{subsec:filter}

The scatter plot of \figurename~\ref{fig:fig_2D_path} shows the data points ([genotype, location, date/time]) 
with respect to two filters, one is time (DAP) and
other is environment (humidity). We created $30$ windows along the filter DAP 
and $5$ windows along the filter humidity, which creates $150$ rectangular interval . 
Each rectangle has a center point.
Initially we did not have any overlapping between two adjacent 
rectangles. We started from $2.5\%$ overlapping between two adjacent   
rectangles by increasing the length of each rectangle 
along both sides from the center point of corresponding rectangle.

\subsection{Partial clusters} 
\label{subsec:partial_cluster}

In the next step to compute partial clusters, the point set from each
rectangular interval was clustered using the algorithm described in 
Section~\ref{sec:Clustering}. The \emph{distance} between any two points in the
set was given by the absolute difference of their trait values. Using the
density-based clustering algorithm, we
generated a set of partial clusters for every rectangular interval. 
Every rectangular interval contains a set of points. 
We calculated standard deviation of phenotypic values for all the points in a rectangle. 
The mean value of all the standard deviations is used as a clustering radius, which is $r=0.7$ for this experiment. 
The density threshold of the clustering is $\rho = 2$, which is fixed in our experiments.

\subsection{Simplicial complex} 
\label{subsec:simplex}

The output of our TDA framework is a simplicial complex constructed using the
overlaps among the set of partial clusters generated from all rectangular
intervals. Recall
that each node represents a partial cluster. In our visual representation of
the complex, the size of each node is scaled to its weighted cardinality (as
defined by the number of core and peripheral points within that cluster).

\subsection{Persistent homology} 
\label{subsec:homology}

As a process to apply homology, which is described in section~\ref{sec:homology}, we generated a set of
overlapping parameters started from $2.5\%$ to $50\%$ with $2.5\%$ interval. For each overlapping
parameter, when a simplicial complex generated, we recorded all the new simplex
information against that overlapping value. This overlapping value is considered as a birth or starting point 
of that new simplex. Similarly, when any existing simplex is replaced by a new 
simplex (i.e. $1$-simplex can be replaced by $2$-simplex when a new cluster overlaps 
two existing clusters of $1$-simplex) then that overlapping value is considered as the 
death or terminating point for the old simplex as well as the birth point for 
 new simplex.
After generating simplicial
complexes for all the overlapping parameters, we got a list of birth and dead overlapping values 
of the simplices. This record is used to generate barcode. 

\begin{figure}[htp!]
\centering
\includegraphics[keepaspectratio=yes, width=\columnwidth]{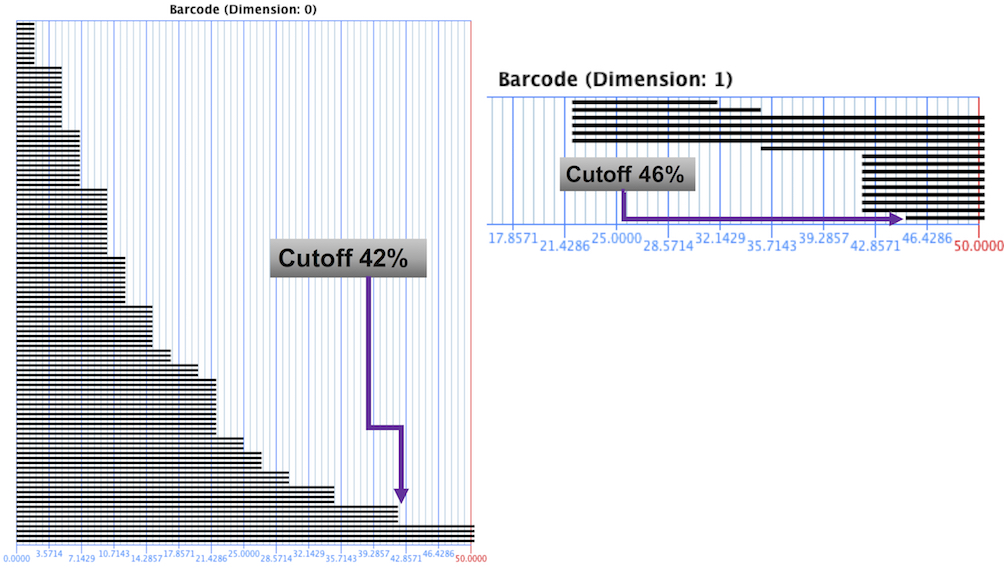}
\caption{Barcode generated from $0\%$ overlapping to $50\%$ overlapping with $2.5\%$ interval. Barcode of 
`dimension $0$' shows the number of connected components and barcode of `dimension $1$' indicates the 
number of holes. Each horizontal bar specifies the life span of a simplex in terms of 
percentage overlapping value. The percentage overlapping value after which there has no any terminal point of 
any simplex is 
considered as the persistent value. Here, the number of connected components are not changing after $42\%$ 
and the number of holes are not changing after $46\%$. These two values are the persistent values 
and we chose the larger one.}
\label{fig:fig_barcode}
\end{figure}

Recall from section~\ref{sec:homology}, barcode for dimension $zero$ indicates the
number of connected components and barcode for dimension $one$ indicates the
number of holes. The barcodes generated from our dataset
(\figurename~\ref{fig:fig_barcode}) shows that the number of connected components did
not change after $42\%$ and the number of holes did not change after $46\%$. We
considered $46\%$ as a constant overlapping value for this double filter function (DAP and Humidity).

\subsection{Topological object}
\label{subsec:topology}

The topological object in \figurename~\ref{fig:fig_path} was generated considering the $46\%$ overlapping. 
The details analysis of this object is in the result section (Section~\ref{sec:dfFltrFnc}).

\end{document}